\documentclass[11pt]{article}

\usepackage[margin=1in]{geometry}
\usepackage{amsmath,amssymb,amsthm,mathtools}
\usepackage{bm}
\usepackage{graphicx}
\usepackage{placeins}
\usepackage{microtype}
\usepackage[hidelinks]{hyperref}

\newcommand{\R}{\mathbb R}
\newcommand{\diag}{\operatorname{diag}}
\newcommand{\spec}{\operatorname{spec}}

\newcommand{\norm}[1]{\left\lVert #1\right\rVert}

\title{Hysteresis and multistability in network spreading with\\ neuronal activity feedback}
\author{Christoffer G. Alexandersen$^{1}$ and Dani S. Bassett$^{1,2,3}$\\[0.75em]
\small $^1$Wu Tsai Institute, Yale University, New Haven, CT, USA\\
\small $^2$Department of Psychology, Yale University, New Haven, CT, USA\\
\small $^3$Department of Biomedical Engineering, Yale University, New Haven, CT, USA}
\date{}

\begin{document}

\maketitle

\begin{abstract}
\noindent Spreading processes on networks often interact with other dynamics on
the same nodes. Neurodegenerative disease provides one example: pathological
proteins spread through anatomical connections, while neuronal activity
influences and is altered by their spread, forming a spreading-activity feedback loop. However, models coupling pathological
protein spreading and neuronal activity have largely focused on linear feedback
between the two processes. Here we show that nonlinear feedback can
fundamentally change the invasion dynamics in a susceptible--infected--susceptible spreading process coupled to a co-evolving activity process. On general weighted networks, the strength and shape of the feedback may generate finite-amplitude invasion thresholds, hysteresis, and endemic multistability.
On regular graphs with homogeneous dynamics, we rule out periodic solutions and show that the degrees of polynomial coupling functions bound the number of endemic states, while monotone couplings require reinforcing feedback for multistability. We test these predictions in simulations of a stochastic spiking neuronal network described by quadratic integrate-and-fire dynamics, where we recover both finite-amplitude invasion thresholds and endemic bistability. These results show that feedback from activity processes can create hysteresis and multistability in network spreading dynamics. In neuroscientific applications, our work suggests that neuronal dynamics may act as a control point in neurodegenerative disease, with even transient changes in activity capable of tipping the brain between health and disease.

\end{abstract}

\newpage

\section{Introduction}

Spreading processes on networks describe a wide range of phenomena including infectious disease, social
contagion, and even the progression of neurodegenerative disease. In the simplest
description, spreading occurs over a fixed network independently of any other
process. In many systems, however, spreading is coupled to a second process that
evolves on the same network. During an epidemic, prevalence may change contact
behavior, while those behavioral changes affect future transmission~\cite{gross_epidemic_2006, frieswijk_mean-field_2022, achterberg_minimal_2023, berry_existence_2025, li_activity_2026}. Disease
burden may deplete resources for prevention or recovery, while resource
availability determines how effectively further spread is controlled
\cite{bottcher_disease-induced_2015}. On online platforms, information diffusion can
create new social connections, while those connections shape subsequent
diffusion \cite{farajtabar_coevolve_2015}. Each of these examples forms a feedback loop in which spreading changes a second dynamical
process, which in turn changes subsequent spreading. Neurodegenerative disease provides another key example, with feedback between pathological protein spreading and neuronal activity.

Many neurodegenerative diseases, such as Alzheimer's disease, involve the accumulation of pathological proteins, which disrupt neuronal function and contribute to disease progression.
These proteins can spread through anatomical connections between brain regions,
so that disease progression can be viewed as a spreading process on the brain's anatomical network
\cite{jucker_self-propagation_2013, goedert_alzheimers_2015}.  Moreover, protein spreading
influences---and is influenced by---neuronal activity, forming a feedback loop.  In Alzheimer's disease, for
example, neuronal activity can promote the release and spread of the
pathological protein tau \cite{pooler_physiological_2013, wu_neuronal_2016}, and
amyloid-$\beta$ plaques can make nearby neurons hyperactive
\cite{busche_clusters_2008}.
Although several network models describe pathological protein spreading on brain networks \cite{raj_network_2012,vogel_spread_2020}, few have considered its coevolution with neuronal dynamics~\cite{alexandersen_multi-scale_2023, alexandersen_neuronal_2024, cabrera-alvarez_multiscale_2024, alexandersen_activity-dependent_2026}. Furthermore, models describing the coevolution of pathology spreading and neuronal dynamics have largely used linear coupling between the two processes, leaving the consequences of nonlinear transmission feedback comparatively unexplored \cite{alexandersen_neuronal_2024, alexandersen_activity-dependent_2026}.

The susceptible--infected--susceptible (SIS) process is a minimal model of
network spreading and provides a natural starting point for studying such
feedback.  In the SIS model, infection spreads between connected nodes, while
infected nodes recover and become susceptible again.  Under common mean-field
approximations, an epidemic state and a disease-free state switch stability through a transcritical
bifurcation when the
transmission-to-clearance ratio crosses a threshold determined by the spectral
radius of the network~\cite{lajmanovich_deterministic_1976, van_mieghem_virus_2009, kiss_mathematics_2017}.
The threshold determines the onset of invasion, beyond which we observe epidemic spreading throughout the network. However, extensions beyond the canonical SIS model such as
nonlinear contact or treatment, higher-order interactions, and adaptive feedback can produce a wider range of dynamical phenomena including
backward bifurcations, hysteresis, discontinuous transitions, additional endemic
states, and sustained oscillations~\cite{van_den_driessche_simple_2000, gross_epidemic_2006, zanette_infection_2008, razvan_global_2012, iacopini_simplicial_2019, huang_backward_2019, wang_bistable_2021, frieswijk_mean-field_2022, achterberg_minimal_2023, berry_existence_2025, li_activity_2026}. Therefore, we may expect similar dynamical phenomena to arise from feedback loops between spreading and neuronal-dynamical processes. 

Motivated by neurodegenerative disease, we couple SIS spreading to general activity dynamics and investigate how feedback between the two processes shapes the bifurcation structure near the invasion threshold. Although our main application is neuronal, we initially keep the activity subsystem general, assuming only that it has an attracting fixed point. The coupling acts locally at each node: the spreading state influences the node's activity, while its activity modulates its outward transmission strength. In contrast to earlier work, we consider general feedback interactions, including nonlinear ones, and determine how they alter the onset and nature of invasion.

We first study invasion near the epidemic threshold on a general weighted
graph. To make the analysis tractable, we use a node-level mean-field approximation of the stochastic SIS process to formulate the coupled network dynamics as a system of ordinary differential equations. Center manifold reduction shows that feedback between spreading and activity produces three
qualitatively different regimes: forward invasion, finite-amplitude invasion
below threshold, and bistability between low- and high-endemic states above
threshold. The bistable regimes produce hysteresis, as increasing and decreasing infection transmission rates can lead the system along different stable branches. To gain intuition in a simpler setting, we consider homogeneous dynamics on regular graphs with a minimal activity model, where the
full nonlinear system reduces to two variables. The homogeneous system cannot sustain periodic
oscillations, and polynomial feedback couplings of
degrees $m$ and $n$ give at most $mn+1$ endemic equilibria, even when the
activity subsystem has a unique stable state.  More generally, monotone
couplings can support multistability only when the feedback between spreading and activity is reinforcing, and linear couplings may only support finite-amplitude invasion and no endemic bistability.
Finally, we test whether finite-amplitude invasion and endemic bistability are present in numerical simulations with biophysical quadratic integrate-and-fire (QIF) neuronal models.
We study both population-level dynamics and networks of individual spiking neurons
coupled to a stochastic SIS process.  Numerical simulations at both levels
reproduce the same multistable dynamics predicted by the
center-manifold calculation and the regular-graph reduction.

This article is organized as follows.  We first study the transcritical bifurcation on a
general graph with the coupled SIS-activity system and identify the center-manifold coefficients that control invasion.
We then study a minimal coupled system on regular graphs to investigate the global nullcline geometry and the
number of homogeneous endemic states.  We subsequently validate these regimes
in biophysical neuronal models.

\section{Invasion near the epidemic threshold on general networks}
\label{sec:continuous_feedback}

We begin by asking how feedback between spreading and activity changes
invasion near the epidemic threshold.  To address this question, we study an SIS
spreading process on a weighted network that coevolves with activity dynamics
at the nodes.  In the underlying stochastic formulation, the SIS states change
through discrete infection and clearance events, whereas the activity variables
evolve continuously according to ordinary differential equations.  We then use
a node-level microscopic Markov-chain approach (MMCA) for the SIS process~\cite{gomez_discrete-time_2010, gomez_nonperturbative_2011, chakrabarti_epidemic_2008}
and replace the stochastic input to activity by its mean-field value.  This
gives a coupled system of ordinary differential equations for the node
infection probabilities and activity, on which we perform the bifurcation
analysis.  Although motivated by pathological protein spreading and neuronal
activity, no particular activity model is assumed.  We compute the epidemic
threshold and perform center manifold reduction to investigate the impact of the spreading-activity feedback on the bifurcation structure near the threshold.
We find forward invasion, finite-amplitude invasion below threshold, and
bistability between endemic states above threshold.

\subsection{Continuous-time infection and activity dynamics}

Let $X_i(t)\in\{0,1\}$, $i=1,\ldots,N$, denote the stochastic infection
state of node $i$, where $X_i=1$ indicates the presence of pathology, and
let $u(t)\in\R^N$ denote the corresponding neuronal activity state.

Conditional on the current infection and activity states, a susceptible
node $i$ becomes infected at rate
\begin{equation}
\label{eq:ctmc_infection_rate}
\lambda_i(X,u)
=
\beta
\sum_{j=1}^N
W_{ij}
\Phi(u_j)
X_j,
\end{equation}
whereas an infected node clears pathology at rate $\zeta$. Here,
$W\geq0$ is an irreducible weighted adjacency matrix, $\beta>0$ is
the baseline transmission rate, and $\zeta>0$ is the clearance rate.  The
smooth function $\Phi:\R\longrightarrow(0,\infty)$ is applied nodewise and
converts source-node activity into an outward
transmission multiplier.  It may be linear, polynomial, sigmoidal, or any
other positive smooth response in the region of state space under
consideration.  Earlier activity--spreading models used a linear dependence of protein transport on neuronal activity
\cite{alexandersen_neuronal_2024, alexandersen_activity-dependent_2026}, with
$\Phi(u_j)=1+\delta u_j$. We use the same symbol $\Phi$ for its componentwise extension to vectors, so that $[\Phi(u)]_j=\Phi(u_j)$ for $u\in\mathbb{R}^N$ and $j=1,\ldots,N$. Whether $\Phi$ denotes the scalar function or its vector extension is therefore clear from its argument.

The neuronal state evolves according to
\begin{equation}
\label{eq:microscopic_activity}
\dot u=\mathcal G(u,X),
\end{equation}
where $\mathcal G:\R^N\times\R^N\longrightarrow\R^N$ is sufficiently
smooth. The infection component is therefore a
continuous-time Markov chain whose transition rates depend on the
continuous activity state. The combined process $(X,u)$ is a
piecewise-deterministic Markov process: infection changes through
discrete jumps, while activity evolves continuously between jumps. To make the calculations more tractable, we introduce simplifying assumptions and study the resulting approximate deterministic system of ordinary differential equations.

Let $x_i(t)=\mathbb E[X_i(t)]$ denote the marginal infection probability.
Applying a node-level mean-field closure to the first-moment equations and
replacing the stochastic activity response by its deterministic mean-field
approximation gives the coupled system
\begin{equation}
\label{eq:full_continuous_system}
\boxed{
\begin{aligned}
\dot x
&=
\beta
(\mathbf 1-x)
\odot
W
\diag\bigl(\Phi(u)\bigr)x
-
\zeta x,
\\
\dot u
&=
\mathcal G(u,x).
\end{aligned}
}
\end{equation}

System~\eqref{eq:full_continuous_system} is a node-level mean-field
approximation that neglects correlations between the infection states of
different nodes and replaces the stochastic input to activity by its mean.
It is expected to be accurate when these correlations and finite-size
fluctuations are weak, as in large, sufficiently well-mixed networks.  It is expected to be less accurate in small or highly structured networks, where
higher-order statistics more strongly influence system dynamics.

\subsection{The linear epidemic threshold}

We study the dynamics near the epidemic threshold of the node-level mean-field SIS-activity system given by Eq.~\eqref{eq:full_continuous_system}.
The epidemic threshold refers to the transcritical bifurcation observed in moment closures of SIS spreading models, where a disease-free state, $x_i = 0$ for all nodes $1 \leq i \leq N$, and an endemic state, $x_i \neq 0$ for some node $i$, switch stability.
To determine the epidemic threshold, we linearize the coupled system about a
disease-free equilibrium.  Suppose that the activity subsystem~\eqref{eq:microscopic_activity} has a
hyperbolically stable equilibrium $a \in \mathbb{R}^N$ in the absence of pathology, so that
$\mathcal G(a,\mathbf 0)=\mathbf 0$ and the activity Jacobian
$J_A=D_u\mathcal G(a,\mathbf 0)$ satisfies $\operatorname{Re}\lambda<0$ for every
$\lambda\in\spec(J_A)$.  The coupled system then has the disease-free
equilibrium $(x,u)=(\mathbf 0,a)$.  We shift the activity variable by defining
$y=u-a$, which maps this equilibrium to $(x,y)=(\mathbf 0,\mathbf 0)$.

To write the linearization compactly, define
$A=\diag\bigl(\Phi(a)\bigr)$, which contains the transmission multipliers at
the disease-free activity state, and $C=D_x\mathcal G(a,\mathbf 0)$, which describes
the linear response of activity to spreading.  In the shifted coordinates,
the Jacobian is
\begin{equation}
\label{eq:disease_free_jacobian}
J_{\mathrm{DF}}(\beta)
=
\begin{pmatrix}
\beta WA-\zeta I & 0\\
C & J_A
\end{pmatrix}.
\end{equation}
The upper-right block vanishes because activity modifies the
transmission of existing pathology but cannot generate pathology when
$x=\mathbf 0$. Since Eq.~\eqref{eq:disease_free_jacobian} is block triangular,
\[
\spec\bigl(J_{\mathrm{DF}}(\beta)\bigr)
=
\spec\bigl(\beta WA-\zeta I\bigr)
\cup
\spec(J_A).
\]
The pathology-to-activity coupling $C$ therefore does not change the
eigenvalues controlling the loss of stability of the disease-free
state. Let $\rho_0=\rho(WA)$, and let $r_0$ and $\ell_0$ be the corresponding right and left
Perron eigenvectors:
\begin{equation}
\label{eq:perron_eigenvectors}
WAr_0=\rho_0r_0,
\qquad
\ell_0^\top WA=\rho_0\ell_0^\top,
\qquad
\ell_0^\top r_0=1.
\end{equation}
The leading infection eigenvalue is
$\lambda_0(\beta)=\beta\rho_0-\zeta$.
Consequently, the disease-free equilibrium loses stability at
\begin{equation}
\label{eq:continuous_epidemic_threshold}
\boxed{
\beta_c
=
\frac{\zeta}{\rho_0}.
}
\end{equation}
Note that no particular form of the activity dynamics is required: its effect on the
threshold enters only through the disease-free activity pattern $a$, as
contained in $A$~\cite{alexandersen_activity-dependent_2026}.  The quantities $J_A$ and $C$ instead determine the activity
response needed for the nonlinear reduction below.

\subsection{Activity response along the critical invasion mode}

To construct the center-manifold reduction near the threshold, we need the right nullspace of
the full Jacobian $J_{\mathrm{DF}}(\beta_c)$.  It is the center eigenspace
tangent to the center manifold, and its activity component describes how
activity changes during invasion.

Because $W$ is irreducible and $A$ has strictly positive diagonal entries,
$WA$ is irreducible and its Perron eigenvalue $\rho_0$ is simple.  For the
infection block $L_c=\beta_cWA-\zeta I$, the identity
$\beta_c\rho_0=\zeta$ therefore gives
$\ker L_c=\operatorname{span}\{r_0\}$, and so the right Perron eigenvector $r_0$ is the infection component of the center eigenspace. Since $J_A$ is invertible, the block
structure in~\eqref{eq:disease_free_jacobian} implies that
$\ker J_{\mathrm{DF}}(\beta_c)$ is also one-dimensional.
Choose its right nullvector as $v_c=(r_0,h)^\top$.  The condition
$J_{\mathrm{DF}}(\beta_c)v_c=\mathbf 0$ gives
$L_cr_0=\mathbf 0$ in the infection block and
$J_Ah+Cr_0=\mathbf 0$ in the activity block, and hence
\begin{equation}
\label{eq:critical_activity_response}
\boxed{
h=-J_A^{-1}Cr_0.
}
\end{equation}

Thus, $h$ is the leading activity response to invasion in the critical
infection pattern $r_0$.  Accordingly, $h_i>0$ means that activity at node
$i$ increases as pathology invades, whereas $h_i<0$ means that it decreases,
to leading order; $h_i=0$ means that there is no leading-order change.  The
vector $Cr_0$ is the immediate activity
perturbation induced by this pattern, while the operator $-J_A^{-1}$
accumulates this forcing over the subsequent activity
relaxation.

\subsection{Center-manifold reduction near the threshold}

Having identified the critical direction in the full state space, we now
reduce the coupled dynamics along it.  The resulting scalar equation describes
how the endemic branch emerges from the threshold and provides the
coefficients used to classify the invasion regimes below.

Let $\mu=\beta-\beta_c$ denote the distance from the epidemic threshold.  At
$\mu=0$, the center manifold through the disease-free equilibrium is tangent
to the critical direction $(r_0,h)^\top$.  For $\beta$ near $\beta_c$, we use
$z=\ell_0^\top x_{\mathrm{cm}}$ as its scalar coordinate, so states on the
manifold satisfy
\begin{equation}
\label{eq:center_manifold_expansion_main}
\begin{aligned}
x_{\mathrm{cm}}(z,\mu)
&=
zr_0+O(z^2,\mu z),
\\
y_{\mathrm{cm}}(z,\mu)
&=
zh+O(z^2,\mu z).
\end{aligned}
\end{equation}
Nearby states approach this form as their stable components decay.  The reduced amplitude equation has the form
\begin{equation}
\label{eq:center_manifold_reduced_main}
\boxed{
\dot z
=
\rho_0\mu z
+
c_2z^2
+
c_3z^3
+
O\!\left(
\mu^2z,
\mu z^2,
z^4
\right).
}
\end{equation}
We use this cubic truncation as the starting point to study the dynamics of the coupled system given by Eq.~\eqref{eq:full_continuous_system} near the epidemic threshold.  The coefficient $\rho_0$ of $\mu z$ is always positive.  By
contrast, $c_2$ and
$c_3$ may either be positive, negative, or zero.  The coefficient $c_2$
determines the generic transcritical orientation, whereas $c_3$ controls the
first curvature and saturation of the endemic branches. Their dependence on the
feedback is described in the next two subsections, with the reduction
summarized in Appendix~\ref{app:center_manifold_reduction}.

\subsection{Quadratic coefficient}

The quadratic coefficient in~\eqref{eq:center_manifold_reduced_main} is
\begin{equation}
\label{eq:c2_explicit}
\boxed{
\begin{aligned}
c_2
={}&
-\zeta\,
\ell_0^\top(r_0\odot r_0)
\\
&+
\beta_c\,
\ell_0^\top
W\mathcal D_1[h]r_0.
\end{aligned}
}
\end{equation}
Here, $\mathcal D_1[h]=\diag\bigl(\Phi'(a)\odot h\bigr)$ denotes the
first-order change in the transmission multipliers along the activity
response $h$.  The calculation is given in
Appendix~\ref{app:center_manifold_reduction}.

The quadratic coefficient $c_2$ determines whether the transcritical bifurcation at $\mu = 0$ is supercritical ($c_2<0)$, subcritical ($c_2>0$), or degenerate ($c_2=0$).
The first term represents the intrinsic SIS saturation. It is negative because
susceptible depletion suppresses further infection, making the transcritical bifurcation supercritical in the absence of the feedback loop. The second term
represents the pathology--activity feedback contribution.
The sign of this contribution is controlled by $\Phi'(a)\odot h$, where $h$
encodes how the activity dynamics respond to pathology, while
$\Phi'(a)$ encodes how activity changes transmission.  When these couplings have
the same sign componentwise, the feedback is reinforcing and the second term is
positive. In this case, the feedback offsets the intrinsic SIS saturation and may drive $c_2$ through
zero, producing the backward orientation required for a finite-amplitude invasion boundary. When their signs are opposite componentwise, the feedback opposes
spreading: the second term is negative and makes $c_2$ more negative.  If the
signs vary across nodes, the overall sign is determined by the Perron-weighted
sum in Eq.~\eqref{eq:c2_explicit}. To find out when, if at all, the endemic branch folds to generate multistability, we calculate the cubic coefficient.

\subsection{Cubic coefficient}

The cubic coefficient in Eq.~\eqref{eq:center_manifold_reduced_main} is
\begin{equation}
\label{eq:c3_explicit}
\boxed{
\begin{aligned}
n_3
={}&
-\beta_c
\left[
r_0\odot WA\xi
+
\xi\odot WAr_0
\right]
\\
&+
\beta_c W
\left[
\mathcal D_1[h]\xi
+
\mathcal D_1[\psi]r_0
+
\frac12\mathcal D_2[h,h]r_0
\right]
\\
&-
\beta_c
r_0\odot W\mathcal D_1[h]r_0,
\\
c_3
={}&
\ell_0^\top n_3.
\end{aligned}
}
\end{equation}
Here, at $\mu=0$, we write $x_{\text{cm}}=zr_0+z^2\xi+O(z^3)$ and
$y_{\text{cm}}=zh+z^2\psi+O(z^3)$, so that $\xi$ and $\psi$ are the second-order
corrections to the infection pattern and the activity response, respectively.
We further define
$\mathcal D_2[h,h]=\diag(\Phi''(a)\odot h\odot h)$.  The calculation is
given in Appendix~\ref{app:center_manifold_reduction}.

The cubic coefficient gives the first indication of whether the endemic branch
can fold. Unlike $c_2$, the sign of $c_3$ cannot generally be inferred from the
coupling directions. It combines susceptible depletion acting on the
curved infection pattern, activity feedback through $\xi$ and $\psi$, the
quadratic part of $\mathcal G$, and the transmission curvature $\Phi''(a)$.
A negative $c_3$ provides cubic saturation, whereas a positive $c_3$
reinforces growth at cubic order.  Its effect on the endemic branches depends
on the sign of $c_2$.

We next determine when $c_2$ and $c_3$ produce a fold of the endemic branch
in the physical region.

\subsection{Local invasion regimes near the epidemic threshold}

A fold of the endemic branch creates a saddle-node bifurcation and can thereby
generate multistability. Retaining terms through cubic order gives
\begin{equation}
\label{eq:cubic_truncation}
\dot z
=
z\left(\rho_0\mu+c_2z+c_3z^2\right).
\end{equation}
For $z\neq0$, equilibria satisfy
$\rho_0\mu+c_2z+c_3z^2=0$. At a saddle-node, two equilibria coalesce, so
$c_2+2c_3z=0$ as well. Hence,
\begin{equation}
\label{eq:saddle_node_amplitude}
z_{\mathrm{SN}}
=
-\frac{c_2}{2c_3},
\qquad
\mu_{\mathrm{SN}}
=
\frac{c_2^2}{4\rho_0c_3}.
\end{equation}
Only $z_{\mathrm{SN}}>0$ corresponds to nonnegative infection probabilities.
Thus, the cubic center-manifold truncation produces a fold in the physical region only when
$c_2$ and $c_3$ have opposite signs.

\paragraph{Backward transition and finite-amplitude invasion.}
A subcritical transcritical bifurcation has $c_2>0$. Its fold lies in the
physical region when $c_3<0$.
The fold then lies below the threshold, with $\mu_{\mathrm{SN}}<0$. The
unstable positive branch from the backward transcritical bifurcation turns at
the saddle-node and continues as a stable endemic branch. For
$\mu_{\mathrm{SN}}<\mu<0$, the disease-free state and a larger endemic state
are stable, while a smaller unstable endemic state separates their basins.
Thus, increasing $\beta$ preserves the disease-free state until $\beta_c$,
whereas decreasing it preserves the endemic state down to
$\beta_{\mathrm{SN}}=\beta_c+\mu_{\mathrm{SN}}$. This gives finite-amplitude
invasion and hysteresis below the threshold.

\paragraph{Forward transition and endemic--endemic multistability.}
A supercritical transcritical bifurcation has $c_2<0$. Its fold lies in the
physical region when $c_3>0$.
The fold then lies above the threshold, with $\mu_{\mathrm{SN}}>0$. The stable
positive branch turns at the saddle-node and continues as an unstable branch.
For $0<\mu<\mu_{\mathrm{SN}}$, the cubic truncation therefore contains a
smaller stable endemic equilibrium and a larger unstable one. This first fold
does not yet give bistability because the disease-free state is unstable.
Endemic--endemic multistability requires a second fold, whose local existence
depends on the fourth-order coefficient $c_4$. Rather than carrying out that
higher-order calculation here, we verify the second fold numerically in the
next subsection.

\paragraph{Degenerate transition.}
The boundary case $c_2=0$ occurs when activity feedback exactly balances the
intrinsic SIS saturation and separates the supercritical and subcritical
orientations. It is nongeneric, and $c_3$ or higher-order terms determine the
leading branch geometry.

\subsection{Numerical validation of the invasion regimes}

We verify in the full deterministic network equations that the physical
folds predicted above can produce finite-amplitude invasion below the
threshold and endemic--endemic bistability above it, rather than occurring
only in the truncated center-manifold equation. We use an Erd\H{o}s--R\'enyi graph with $20$ nodes, where each edge is included with probability 0.2. Its adjacency
matrix is normalized so that $\rho(W)=1$, and we set the SIS clearance $\zeta=1$.  The activity
dynamics are kept minimal with a linear spreading-to-activity coupling,
$\dot u_i=-u_i+x_i$, and we vary only the activity-to-spreading coupling
$\Phi$.

The linear coupling $\Phi(u)=1+u/2$ produces monostable forward invasion, while the quadratic polynomials $\Phi(u)=(1+u)^2$ and $\Phi(u)=1+3u/5+5u^2/2$ produce finite-amplitude invasion below the epidemic threshold $\beta<\beta_c$ and endemic--endemic bistability above the threshold $\beta>\beta_c$.
For each choice of $\Phi$, we compute the numerical bifurcation diagram
of the full $40$-dimensional system as $\beta$ is varied. 
Each endemic
equilibrium is represented by its infection amplitude $z=\ell_0^\top x$.
The numerical continuation and stability calculation is described in
Appendix~\ref{app:irregular_network_numerics}.
Figure~\ref{fig:irregular_network_validation}(a--c) plots the resulting
branches together with the local center-manifold prediction.  The dash-dotted
lines mark the values of $\beta$ used for direct
integration of the full system in panels (d--f).  The center-manifold reduction approximates
the branches well close to the bifurcation.

We then integrate the full equations through time from lower and higher initial
conditions. The trajectories confirm the stability and initial-condition
dependence of the branches in
Figure~\ref{fig:irregular_network_validation}(d--f).

\begin{figure}[!htb]
\centering
\includegraphics[width=\textwidth]{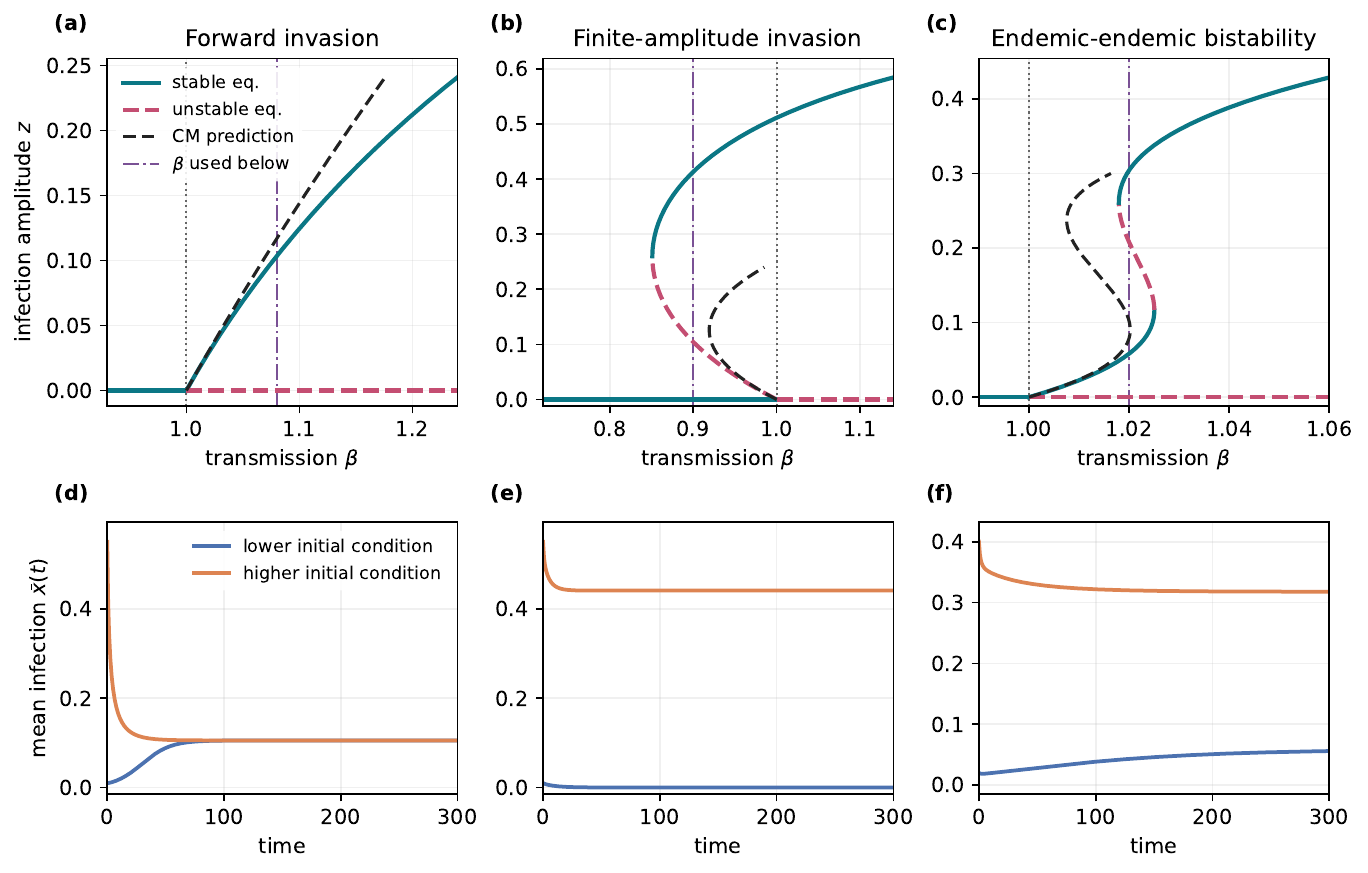}
\caption{Numerical bifurcations on an Erd\H{o}s--R\'enyi graph
$G(20,0.2)$. (a--c) Endemic branches for $\Phi(u)=1+u/2$,
$\Phi(u)=(1+u)^2$, and $\Phi(u)=1+3u/5+5u^2/2$, respectively. Black dashed
curves show the center-manifold predictions. Solid blue and dashed red branches are stable and
unstable, black dotted vertical lines mark $\beta_c=1$, and dash-dotted purple lines mark the values
of $\beta$ used in (d--f). (d--f) Mean infection from lower and higher initial
conditions at $\beta=1.08$, $0.90$, and $1.02$, respectively. In (d,e), the
uniform initial states have $x_i(0)=u_i(0)=0.01$ or $0.55$; in (f), they have
$x_i(0)=u_i(0)=0.02$ or $0.40$.}
\label{fig:irregular_network_validation}
\end{figure}

\section{Endemic multistability on regular graphs}
\label{sec:regular_polynomial_feedback}

The center-manifold reduction shows how endemic branches emerge near the
epidemic threshold, but it cannot determine how many endemic states exist in
the full nonlinear system or what happens farther from the threshold. To
address these questions in a simpler setting, we now restrict the system to homogeneous dynamics on
a regular graph. 
For general smooth coupling functions, we show that the number of endemic equilibria is even below the epidemic threshold and odd above it, and we exclude periodic behavior. We further show that multistability is impossible for opposite-sign monotone coupling. For polynomial couplings, we find a bound on the number of endemic equilibria, and for sigmoidal couplings, we identify bistable regimes.

\subsection{Homogeneous mean-field reduction}

Suppose that $W$ represents a regular graph in which every node has weighted
degree $\rho$, so that $W\mathbf 1=\rho\mathbf 1$.  We assume that all edge
weights are nonnegative and that the graph is connected, or strongly connected
if its edges are directed. Equivalently, $W$ is an irreducible nonnegative
matrix. The Perron--Frobenius theorem
then implies that $\rho$ is the spectral radius of $W$.
A homogeneous initial condition remains homogeneous: if every
node starts with the same infection probability $x$ and activity deviation
$y$, all nodes continue to have the same values. We therefore restrict the
following analysis to these homogeneous dynamics, which are described by one
infection variable and one activity variable.

As before, $\Phi$ denotes the activity-to-transmission coupling. We write
$H(x)$ for the pathology-to-activity coupling and require that $H(0)=0$.
The homogeneous dynamics are
\begin{equation}
\label{eq:regular_polynomial_system}
\boxed{
\begin{aligned}
\dot x
&=
x\left[
\beta\rho(1-x)\Phi(a+y)-\zeta
\right],
\\
\dot y
&=
-y+H(x).
\end{aligned}
}
\end{equation}
Because $H(0)=0$, the disease-free
equilibrium is $(x,y)=(0,0)$, with invasion threshold
$\beta_c=\zeta/[\rho\Phi(a)]$.

\subsection{General constraints on the homogeneous dynamics}

The homogeneous system
constrains the number and stability of its endemic equilibria and rules out
sustained oscillations. Let $\alpha=\zeta/(\beta\rho)$. At an
endemic equilibrium of the homogeneous system~\eqref{eq:regular_polynomial_system}, the activity nullcline gives $y=H(x)$, so the infection
coordinate is a root in $(0,1)$ of
\begin{equation}
\label{eq:regular_equilibrium_polynomial}
F(x)
:=
(1-x)\Phi\bigl(a+H(x)\bigr)-\alpha.
\end{equation}
The endpoint signs and the direction of each crossing determine how many
equilibria can occur on either side of the threshold and which of them are
stable. Because $F(0)=\Phi(a)-\alpha$ is positive above $\beta_c$ and negative
below it, while $F(1)=-\alpha<0$, the number of simple endemic roots is odd
above $\beta_c$ and even below it. At a simple endemic root $x_*$, the trace
and determinant of the Jacobian of the homogeneous system in
Eq.~\eqref{eq:regular_polynomial_system} satisfy
\begin{equation}
\label{eq:regular_jacobian_signs}
\operatorname{tr}J
=
-\beta\rho x_*\Phi(a+H(x_*))-1<0,
\qquad
\det J
=
-\beta\rho x_*F'(x_*).
\end{equation}
Hence downward crossings of $F$ are stable and upward crossings are saddles.
Simple endemic equilibria alternate in stability, and the endemic equilibrium
with the highest pathology level is always stable.

The homogeneous system also cannot support sustained oscillations. Choosing
the Dulac function $B(x,y)=1/x$ on $x>0$, the Bendixson--Dulac criterion
excludes periodic orbits in the physical region using the assumed positivity
of $\Phi$.

The equilibrium function also gives a global sign rule for monotone
couplings. Differentiating Eq.~\eqref{eq:regular_equilibrium_polynomial} gives
\begin{equation}
\label{eq:regular_feedback_derivative}
F'(x)
=
-\Phi(a+H(x))
+(1-x)\Phi'(a+H(x))H'(x).
\end{equation}
Thus, $F$ turns upwards precisely when
\begin{equation}
\label{eq:regular_feedback_turning_condition}
(1-x)\Phi'(a+H(x))H'(x)
>
\Phi(a+H(x)).
\end{equation}
If one coupling is non-decreasing and the other is non-increasing, then
$\Phi'(a+H(x))H'(x)\leq0$, so the inequality cannot hold. Hence, $F$ is
strictly decreasing, giving no endemic equilibrium at or below $\beta_c$ and
exactly one stable endemic equilibrium above it. If the couplings are
monotone in the same direction, the inequality may hold when the feedback is
sufficiently strong. Same-sign coupling therefore permits
multistability, but does not guarantee it. For opposite-sign coupling,
however, multistability is ruled out.

The Bendixson--Dulac calculation is given in
Appendix~\ref{app:regular_periodic_exclusion}.

\subsection{Polynomial bounds on endemic multistability}

Polynomial coupling functions provide a natural starting point for studying
nonlinear feedback and give an explicit bound on the number of equilibria. If
$\Phi$ and $H$ have degrees $m$ and $n$, respectively, then the equilibrium
function $F$ in Eq.~\eqref{eq:regular_equilibrium_polynomial} has degree at
most $mn+1$. Hence
\begin{equation}
\label{eq:regular_root_bound}
N_{\mathrm{endemic}}\leq mn+1.
\end{equation}
If both couplings are linear, then $m=n=1$ and the bound allows at most two
endemic equilibria. However, we know that the number of endemic equilibria is odd above the threshold, so there is exactly one stable endemic
equilibrium. Below it, we have an even number of endemic equilibria, so there are either no endemic equilibria or a
saddle--stable pair. Therefore, reinforcing linear coupling can produce
disease-free--endemic bistability, but not endemic--endemic bistability. Higher
polynomial degrees, however, permit additional endemic states.

We next show that higher-degree polynomials can produce more stable endemic equilibria. We fix the linear activity dynamics $\dot y=-y+x$, set
$a=0$ and $\rho=\zeta=1$, and vary only $\Phi$. The transmission functions and
their exact factorizations are given in
Appendix~\ref{app:regular_exact_examples}.

Figure~\ref{fig:regular_graph_multistability}(a--c) shows the phase planes for
quadratic, cubic, and quartic transmission functions. They produce two stable
endemic states above the threshold, the disease-free state together with two
stable endemic states below the threshold, and three stable endemic states
above the threshold, respectively. Direct integrations in panels (d--f)
confirm that the initial condition selects among these attractors.

\begin{figure}[!htb]
\centering
\includegraphics[width=\textwidth]{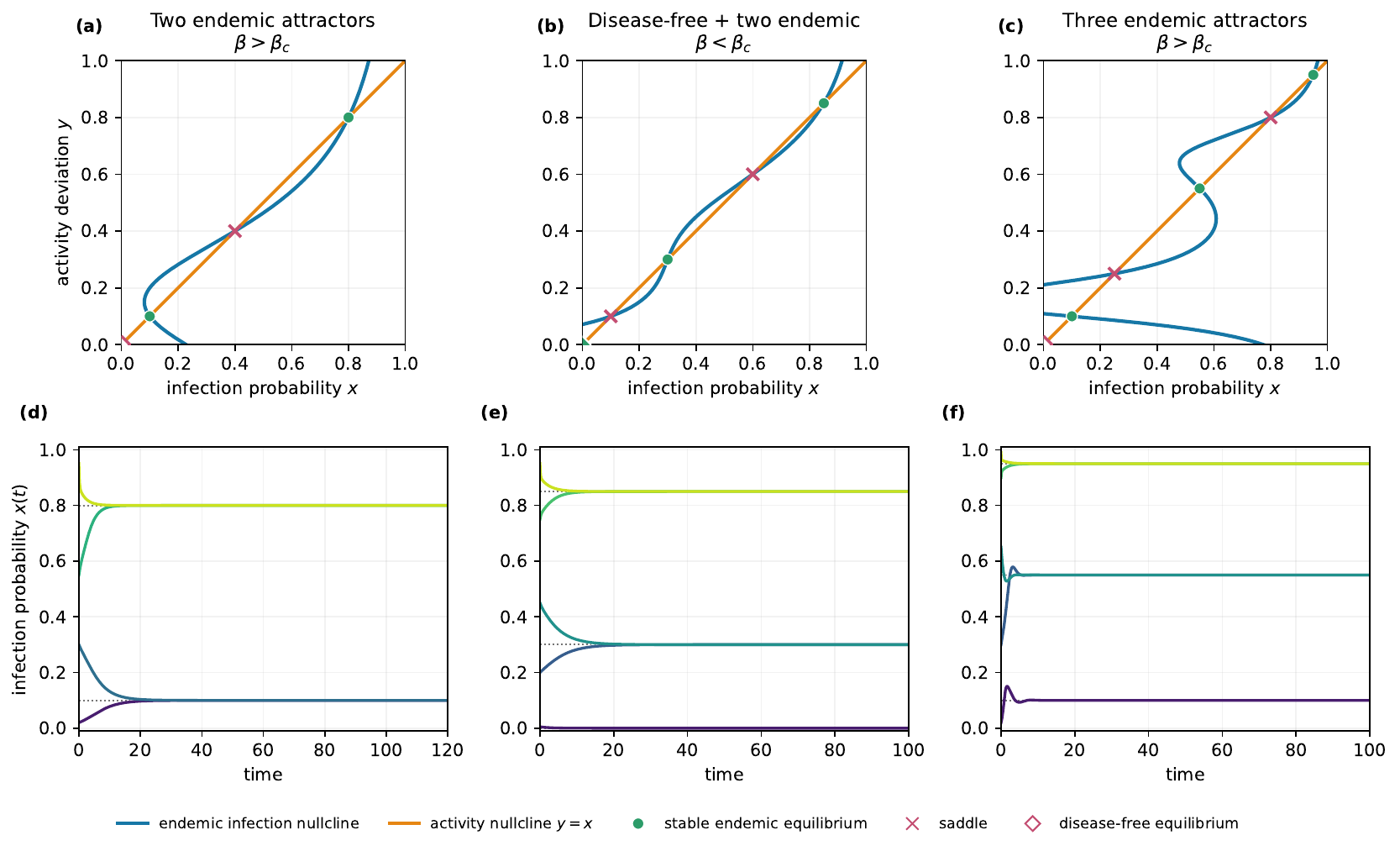}
\caption{Multistability generated by polynomial activity-to-transmission
coupling with the linear activity dynamics $\dot y=-y+x$. (a--c) Phase planes
for quadratic, cubic, and quartic transmission functions. (d--f) The
corresponding direct integrations from different homogeneous initial
conditions with $y(0)=x(0)$. Dotted horizontal lines mark the stable
equilibria. Filled disease-free diamonds are stable and open diamonds are
unstable.}
\label{fig:regular_graph_multistability}
\end{figure}

\subsection{Multistability with sigmoidal couplings}
\label{sec:regular_sigmoidal_feedback}

Polynomials give a general description of nonlinear feedback, but biological
effects, including those in neurodegeneration, often saturate. The global sign
rule above already shows that opposite-sign monotone couplings, such as sigmoids, cannot produce
multistability. We therefore test whether same-sign sigmoidal couplings can
produce multistability and, if so, how many stable equilibria they can support.

Let $S_x$ and $S_u$ be increasing logistic functions normalized between zero
and one. We take
\[
\Phi(u)=1+\delta S_u(u),
\qquad
H(x)=\eta S_x(x),
\]
where the signed gains $\delta$ and $\eta$ set the strength and direction of
the activity-to-spreading and spreading-to-activity couplings.

\begin{figure}[!htb]
\centering
\includegraphics[width=\textwidth]{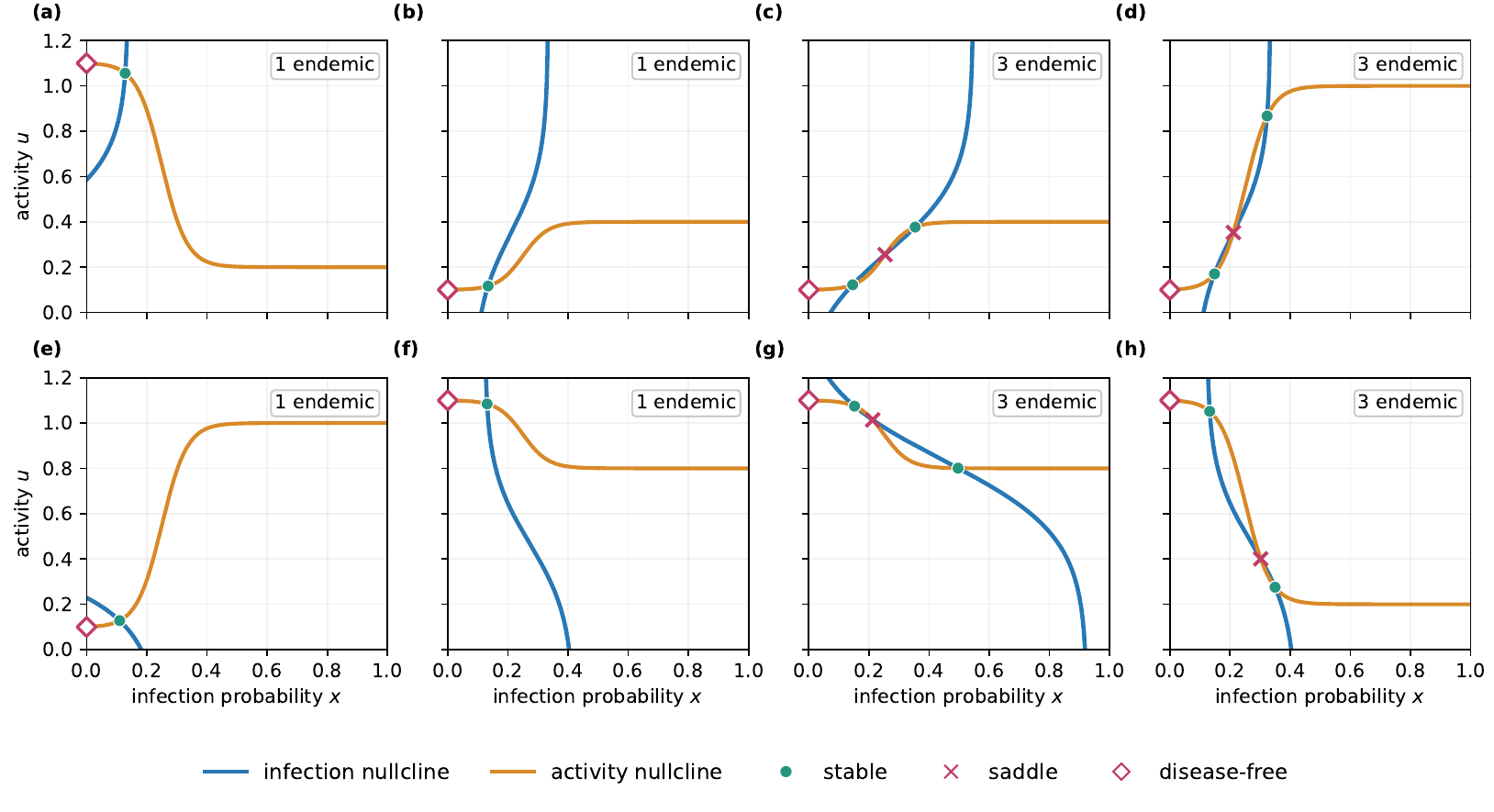}
\caption{Nullclines and endemic equilibria for sigmoidal coupling.
(a,e) Opposite signs with $(\delta,\eta)=(0.90,-0.90)$ and
$(-0.90,0.90)$. (b,f) Same signs with $(\delta,\eta)=(0.30,0.30)$ and
$(-0.30,-0.30)$. (c,g) Increasing $|\delta|$ to $0.90$ produces
multistability; (d,h) increasing $|\eta|$ does the same. Blue and orange curves
show the infection and activity nullclines. Filled circles are stable endemic
equilibria, crosses are saddles, and open diamonds are unstable disease-free
equilibria. All panels use $\beta/\beta_c=1.15$; the two sigmoids have
steepnesses $24$ and $6$ and
transition points $x=0.25$ and $u=0.40$, respectively.}
\label{fig:sigmoidal_feedback_nullclines}
\end{figure}

Figure~\ref{fig:sigmoidal_feedback_nullclines} compares the four sign
combinations. Opposite-sign couplings give one stable endemic equilibrium.
For same-sign couplings, gains of magnitude $0.30$ also give one endemic
equilibrium, whereas increasing the magnitude of either gain to $0.90$
produces two stable endemic equilibria separated by a saddle. Thus, same-sign
sigmoidal coupling can produce multistability when the feedback is sufficiently
strong.

\section{Invasion and multistability in QIF neuronal models}
\label{sec:qif_application}

We next test the predicted invasion regimes in QIF neuronal models at two
spatial scales. First, we use a regular network of mean-field QIF populations,
where each node represents a neuronal population, to calculate the bifurcation
structure. We then model a neural circuit in which each node is an individual
spiking QIF neuron coupled to stochastic SIS spreading.

\subsection{Bifurcation structure in mean-field QIF populations}
\label{sec:qif_macroscopic_validation}

We model a brain network in which each node represents the
population-averaged dynamics of many neurons, using the exact mean-field
equations for a heterogeneous QIF population \cite{montbrio_macroscopic_2015}. The
populations are coupled both to spreading and to each other. Along each neural
edge $j\to i$, all neurons in population $j$ project uniformly to all neurons
in population $i$, giving the common input $KW_{ij}r_j$ while preserving the
exact QIF reduction \cite{avitabile_cross-scale_2022}. As before, $H(x)$ describes
infection-induced changes in excitability and $\Phi(r)$ the effect of firing on
transmission.
\begin{equation}
\label{eq:qif_graph_system}
\boxed{
\begin{aligned}
\dot x_i
&=
\beta
(1-x_i)
\sum_j W_{ij}\Phi(r_j)x_j-x_i,
\\
\varepsilon\dot r_i
&=
\frac{\Delta}{\pi}+2r_iv_i,
\\
\varepsilon\dot v_i
&=
v_i^2+\bar\eta+H(x_i)+J_{\mathrm{loc}}r_i
+K\sum_jW_{ij}r_j-(\pi r_i)^2.
\end{aligned}
}
\end{equation}
Here $x_i$ is the infection probability, while $r_i$ and $v_i$ are the firing
rate and mean membrane potential of population $i$. The parameter
$\varepsilon$ is the ratio between neuronal and spreading timescales, $\Delta$
is the half-width of the Lorentzian excitability distribution, and $\bar\eta$
is its mean. The gains $J_{\mathrm{loc}}$ and $K$ set the neuronal coupling within and between
populations, respectively. Time is measured in units of the mean recovery
time, so the SIS recovery rate is one.

To calculate the bifurcation structure, we restrict the analysis to
homogeneous initial conditions on a regular graph. All nodes then remain
identical, so their equilibrium branches can be followed in a homogeneous
system. We use a normalized 40-node ring network with four connections of weight $1/4$ per node. On the homogeneous
system, the interpopulation input reduces to $Kr$. Both couplings are bounded
sigmoids with steepness $20$. Let $r_0$ denote the stationary firing rate
without infection. The
pathology sigmoid $S_x$ is centered at $x=0.25$ and is therefore weak when
pathology is rare, while the firing sigmoid $S_r$ is centered at $r=r_0$ and
detects changes from baseline firing. Both vanish at zero and approach one. We
define
\begin{equation}
\label{eq:qif_sigmoidal_couplings}
H(x)=\kappa S_x(x),
\qquad
\Phi(r)=1+\delta S_r(r).
\end{equation}
Here, $\kappa$ and $\delta$ are the spreading-to-activity and
activity-to-spreading gains, respectively. Their signs set the two coupling
directions, and $\Phi(0)=1$. Consequently, the epidemic
threshold is $\beta_c=1/\Phi(r_0)$.

For $J_{\mathrm{loc}}+K\leq0$, the homogeneous QIF system has a unique positive firing rate
$r_*(x)$ for every fixed infection level
(Appendix~\ref{app:qif_monostability}).  The sign rule in Sec.~\ref{sec:regular_polynomial_feedback} permits  multistability when the feedback between the neural dynamics and spreading is reinforcing and
sufficiently strong. Since the firing rate is unique at each infection level,
this multistability must come from the feedback loop. We vary $\kappa$ and
$\delta$ to test for finite-amplitude invasion and endemic--endemic
bistability.

\begin{figure}[!htb]
\centering
\includegraphics[width=\textwidth]{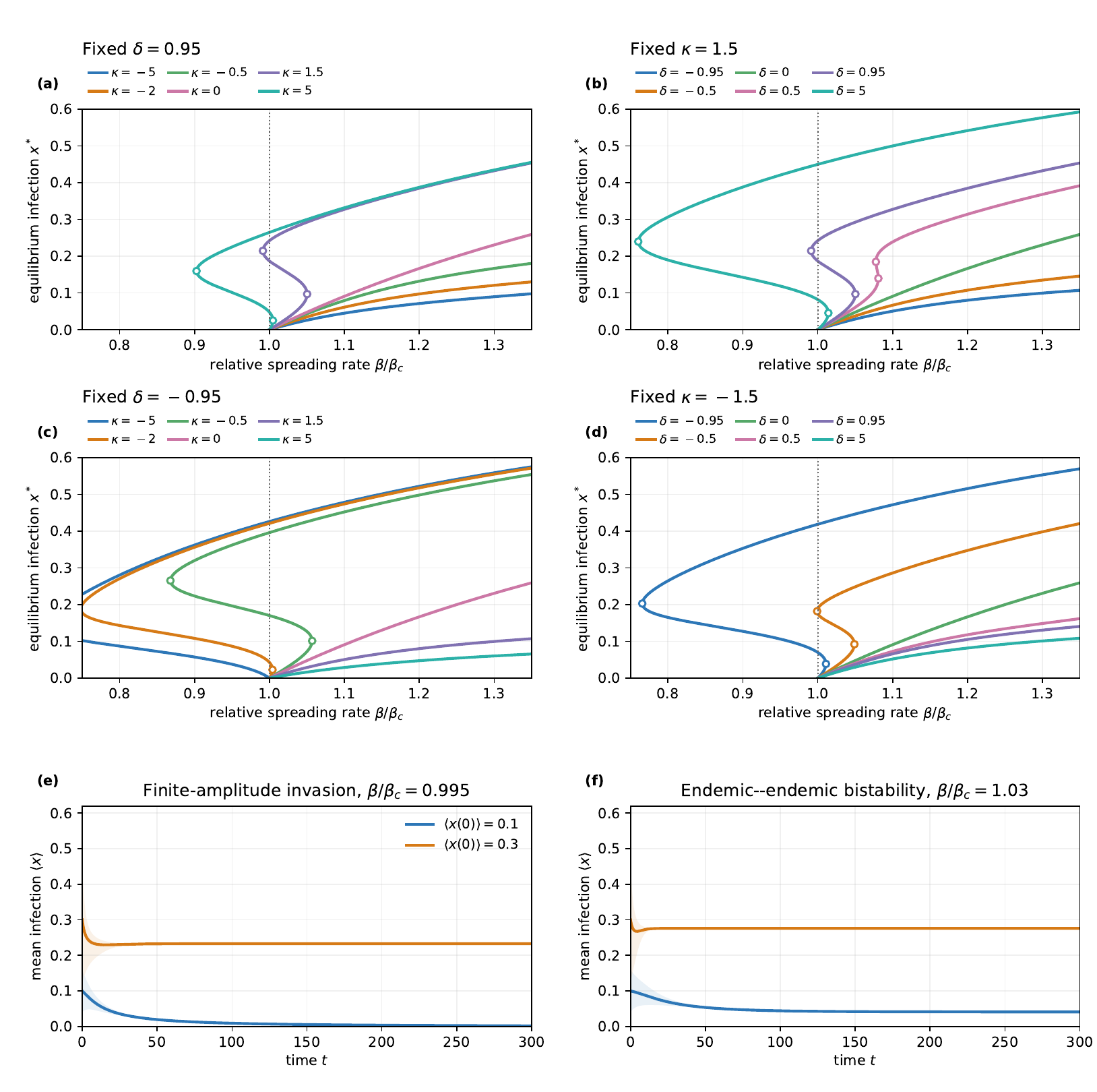}
\caption{Effect of the coupling gains in the mean-field QIF populations.
(a,b) Equilibrium branches as $\kappa$ or $\delta$ changes sign with the other
gain fixed and positive. (c,d) Corresponding sweeps with the other gain fixed
and negative. Circles mark saddle-node bifurcations and dotted lines mark the
epidemic threshold. (e) Finite-amplitude invasion below the threshold.
(f) Endemic--endemic bistability above it. Simulations use
$(J_{\mathrm{loc}},K,\kappa,\delta)=(-2,1,1.50,0.95)$ on a 40-node ring, with mean initial
infection levels $0.10$ and $0.30$. Curves show the network mean and shading
shows the range over nodes.}
\label{fig:qif_feedback_regimes}
\end{figure}

On the homogeneous ring, the equilibrium firing rate at infection level $x$
is $r_*(x)$. Since the recovery rate and spectral radius are both one, the
endemic equilibria are the roots of
\begin{equation}
\label{eq:qif_equilibrium_function}
F(x;\beta)
=
(1-x)\Phi(r_*(x))-\frac{1}{\beta}.
\end{equation}
We set $J_{\mathrm{loc}}=-2$, $K=1$, $\bar\eta=0.25$, $\Delta=0.1$, and
$\varepsilon=0.05$. These values keep the homogeneous QIF system monostable,
giving the disease-free firing rate $r_0\simeq0.121$. We then vary $\kappa$ and $\delta$ to obtain
the equilibrium branches in
Figure~\ref{fig:qif_feedback_regimes}(a--d).

Panels (a,c) vary $\kappa$ at fixed $\delta=\pm0.95$, while panels (b,d) vary
$\delta$ at fixed $\kappa=\pm1.50$. Opposite-sign couplings give a single
endemic branch, whereas sufficiently strong same-sign couplings produce two
saddle-node bifurcations. For $(\kappa,\delta)=(1.50,0.95)$, these occur at
$\beta/\beta_c\simeq0.991$ and $1.050$. Direct simulations with the same
parameters verify finite-amplitude invasion below the threshold in panel (e)
and endemic--endemic bistability above it in panel (f).

\subsection{Multistability in stochastic spiking networks}
\label{sec:qif_microscopic_validation}

We now model a stochastic neural circuit in which each node is an individual
spiking QIF neuron coupled to stochastic SIS spreading. The QIF neuron is
equivalent to the theta-neuron model under a change of variables. In this
form, the membrane state of neuron $i$ is represented by a phase $\theta_i$
that moves around a circle, and a crossing of $\theta_i=\pi$ is recorded as a
spike. The variable $X_i(t)\in\{0,1\}$ is now a susceptible/infected state
rather than an infection probability. The neural dynamics obey
\begin{equation}
\label{eq:microscopic_qif_theta}
\varepsilon\dot\theta_i
=
1-\cos\theta_i
+(1+\cos\theta_i)
\left[
\eta_i+\kappa X_i+J\sum_jW_{ij}s_j+I_i^{\rm ext}(t)
\right].
\end{equation}
Here the excitabilities $\eta_i$ are chosen to represent a Lorentzian distribution with mean
$\bar\eta$ and half-width $\Delta$, as detailed in Appendix~\ref{app:microscopic_numerical_method}. The parameter $\varepsilon$ sets the neuronal timescale, $J$ is the
synaptic coupling strength, and $s_j(t)$ is a time-dependent measure of the
recent firing of neuron $j$, whose evolution is defined below by
Eq.~\eqref{eq:microscopic_spike_filter}.
Infection changes the current through $\kappa X_i$, while
$I_i^{\rm ext}(t)$ is an externally applied neuronal input.
The recent spike history of neuron $i$ is represented by the exponential
filter
\begin{equation}
\label{eq:microscopic_spike_filter}
\tau_s\dot s_i
=
-s_i+\varepsilon\sum_k\delta_{\rm D}(t-t_i^k),
\end{equation}
where $\tau_s$ is the filter timescale and $t_i^k$ are the spike times. 

When
node $j$ is infected, its own recent
firing changes its outward transmission through the same sigmoidal form
$S_r$ used in Eq.~\eqref{eq:qif_sigmoidal_couplings}. Conditional on the
current neural state, the SIS transitions are
\begin{equation}
\label{eq:microscopic_sis_hazards}
\begin{aligned}
X_i:0\longrightarrow1
&\quad\text{at rate}\quad
\lambda_i(t)
=\frac{\beta}{\tau_X}
\sum_jW_{ij}\bigl(1+\delta S_r(s_j(t))\bigr)X_j(t),
\\
X_i:1\longrightarrow0
&\quad\text{at rate}\quad
\frac{\zeta}{\tau_X}.
\end{aligned}
\end{equation}
Here $\beta$ and $\zeta$ set the transmission and recovery rates,
$\tau_X$ is the spreading timescale, and $\delta$ is the
activity-to-spreading gain. We simulate the coupled system by alternating a
QIF integration step with exact Gillespie sampling of the SIS
events~\cite{bauer_multiscale_2018, gillespie_exact_1977}. The full algorithm is given in
Appendix~\ref{app:microscopic_simulation}.

To test whether the same-sign multistability found in the mean-field QIF model
(Figure~\ref{fig:qif_feedback_regimes}) persists in a spiking network, we take
$\kappa>0$ and $\delta>0$. We draw the positions of $N=1000$ nodes independently from a uniform
distribution over the unit square and connect every pair separated by less
than $0.155$. The resulting connected graph has $32\,893$
undirected edges and mean degree $65.8$. We normalize the edge weights by setting
$W=A/\rho(A)$, so that $\rho(W)=1$. The neuronal parameters are
$\bar\eta=-3.25$, $\Delta=0.1$, $J=6$, and $\kappa=3$, while the transmission
gain is $\delta=12$. The firing sigmoid is centered at $s=0.1$ and has
steepness $2$.
Neuronal activity evolves over
milliseconds to seconds, whereas pathological spreading occurs over months to
years~\cite{alexandersen_neuronal_2024, alexandersen_network_2026}. We therefore set
$\tau_X=100$, $\varepsilon=0.02$, and $\tau_s=0.05$, so that many spikes occur
between infection and recovery events.

\begin{figure}[!htb]
\centering
\includegraphics[width=\textwidth]{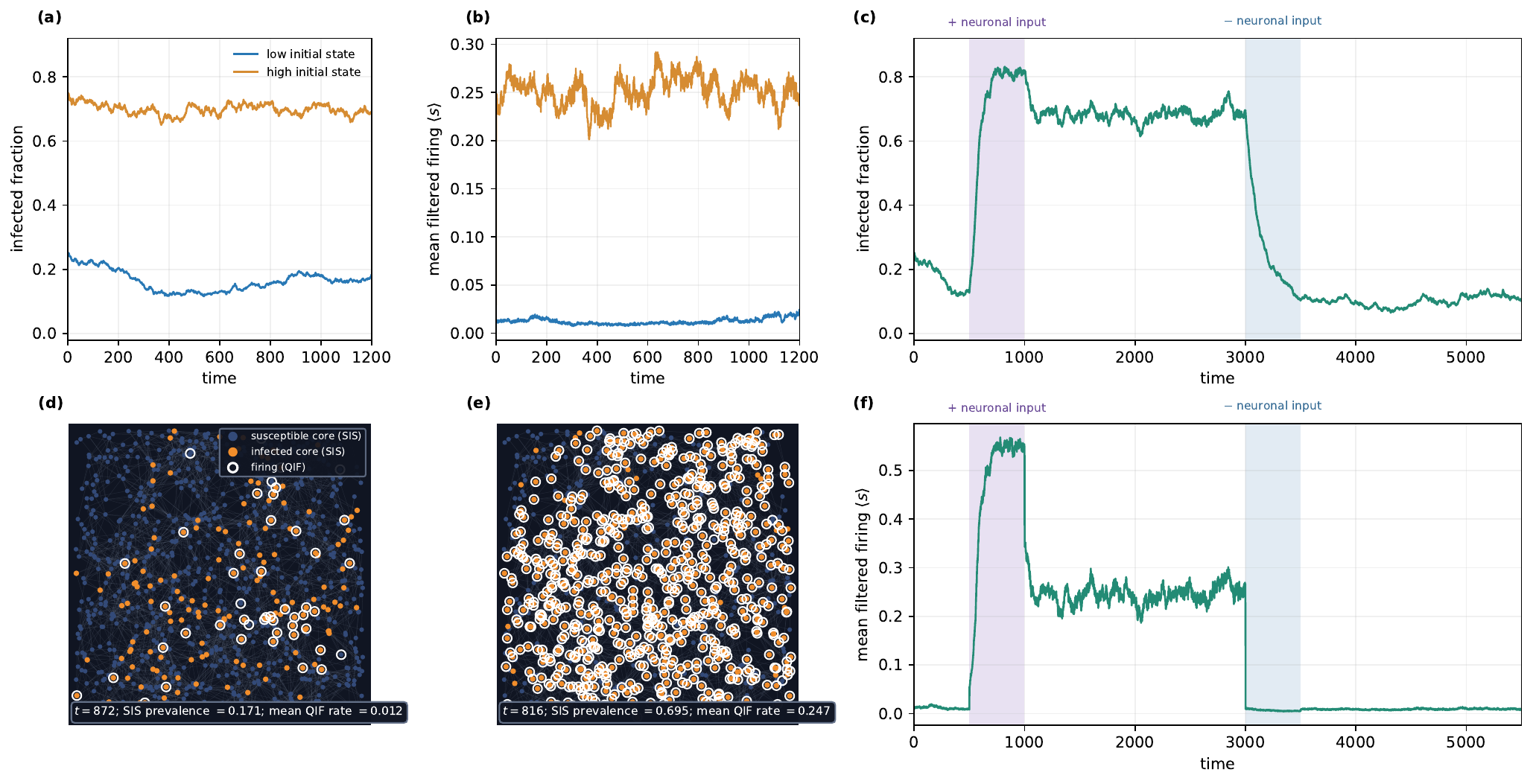}
\caption{Endemic--endemic bistability in the stochastic SIS--QIF network.
(a,b) Infection prevalence and mean filtered firing from low and high initial
infection levels. (c,f) Positive (purple) and negative (blue) neuronal inputs
move the network between the two endemic regimes. (d,e) Representative
spatial states; blue and orange nodes are susceptible and infected, and white
rings mark spikes. The network has $N=1000$ nodes, connection radius $0.155$,
and mean degree $65.8$. Parameters are $\beta=0.90$, $\bar\eta=-3.25$,
$\Delta=0.1$, $J=6$, $\kappa=3$, $\delta=12$, $\tau_X=100$,
$\varepsilon=0.02$, and $\tau_s=0.05$. Initial infected fractions are $0.25$
and $0.75$, and the transmission sigmoid has steepness $2$ and midpoint
$0.1$. Inputs have amplitude $1$ and act over $500\leq t<1000$ and
$3000\leq t<3500$.}
\label{fig:microscopic_qif_stochastic_regimes}
\end{figure}

We first test whether the endemic--endemic bistability found in the mean-field
QIF populations is also present in the stochastic spiking network. At
$\beta=0.90$, initial infected fractions $0.25$ and $0.75$ remain near mean
prevalences $0.175$ and $0.697$, respectively, showing distinct low- and
high-endemic regimes
(Figure~\ref{fig:microscopic_qif_stochastic_regimes}(a,b)). We then ask whether
neuronal input can move the network between these regimes. Starting from the
lower regime, we set $I_i^{\rm ext}(t)=1$ at every node for $500\leq t<1000$
and $I_i^{\rm ext}(t)=-1$ for $3000\leq t<3500$, with zero input otherwise.
The first input moves the stochastic SIS process to the high-endemic regime,
while the second returns it to the low-endemic regime
(Figure~\ref{fig:microscopic_qif_stochastic_regimes}(c,f)). Both regimes remain
after the corresponding input is removed. Thus, the stochastic spiking
network exhibits endemic--endemic bistability, and transient neuronal input
can move it between the two endemic regimes. A realization of the stochastic simulation with external input is shown in Supplementary
Video S1.

\begin{figure}[!htb]
\centering
\includegraphics[width=\textwidth]{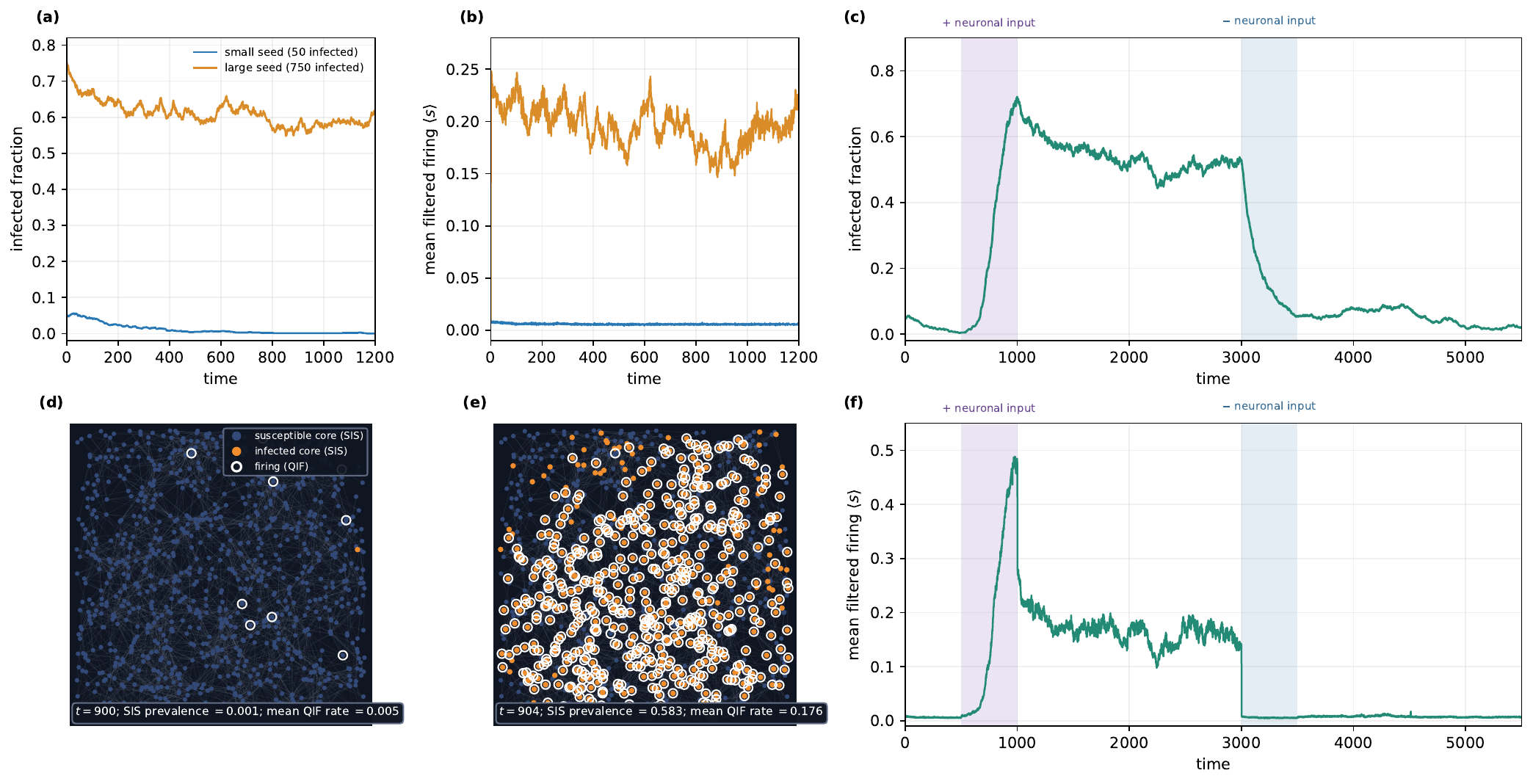}
\caption{Finite-amplitude invasion in the stochastic SIS--QIF network at
$\beta=0.76$. (a,b) A 50-node infection seed becomes extinct, whereas a
750-node seed reaches a high-endemic regime. (c,f) Starting from the same
50-node seed, positive (purple) and negative (blue) neuronal inputs move the
network to the high-endemic regime and back to low infection and activity.
(d,e) Representative disease-free and high-endemic spatial states. The
network, remaining parameters, and input intervals are as in
Figure~\ref{fig:microscopic_qif_stochastic_regimes}.}
\label{fig:microscopic_qif_subthreshold_bistability}
\end{figure}

To test for finite-amplitude invasion at lower transmission, we reduce
$\beta$ from $0.90$ to $0.76$ in
Figure~\ref{fig:microscopic_qif_subthreshold_bistability}. With identical neuronal initial
conditions, a seed of $50$ infected nodes reaches disease-free absorption,
whereas a seed of $750$ nodes reaches a high-endemic metastable regime.
Thus, the small infection seed dies out, but infection persists when its initial
level is sufficiently large. We then ask whether neuronal input can move the
network across this invasion boundary. Starting from the 50-node seed and the
same neuronal initial condition, we set $I_i^{\rm ext}=1$ at every node for
$500\leq t<1000$ and
$I_i^{\rm ext}=-1$ for $3000\leq t<3500$, with zero input otherwise. The first
input moves the stochastic SIS process to the high-endemic regime, while the
second returns it to the low-infection state
(Figure~\ref{fig:microscopic_qif_subthreshold_bistability}(c,f)). Both states
remain after the corresponding input is removed. Thus, the spiking neural
network exhibits finite-amplitude invasion, and transient neuronal input can
move it across the invasion boundary. A realization of the stochastic simulation with external input is shown in Supplementary
Video S2.

To distinguish this behavior from a long transient, we repeat the input
experiment in Appendix~\ref{app:microscopic_opposite_sign_control} after
reversing the sign of $\kappa$, so that the two coupling directions have
opposite signs. Even with a five-times stronger input, infection begins to
decay once the input is removed and eventually reaches absorption
(Figure~\ref{fig:microscopic_qif_opposite_sign_control}). This comparison
therefore supports interpreting the high-endemic state in
Figure~\ref{fig:microscopic_qif_subthreshold_bistability} as the result of
crossing a finite-amplitude invasion boundary.

\section{Discussion}

We have studied a spreading process which co-evolves with a general activity
process on the same network. The feedback loop between activity and spreading
determines
whether invasion is continuous, requires a finite perturbation, or leads to
several endemic states. These dynamics can occur even when the activity
subsystem has a unique stable state, in which case the multistability is generated by the
feedback loop itself. Numerical simulations of biophysical neuronal models
recover the same dynamical behavior, both in mean-field QIF populations and
in networks of individual spiking QIF neurons coupled to stochastic SIS
spreading.

\subsection{Activity-driven networks and other routes to multistability}

Activity-driven networks provide the closest comparison to the present work.
In the original framework, nodes create temporary contacts at heterogeneous
rates, allowing contagion to spread over a time-varying network
\cite{perra_activity_2012}. Subsequent work established how activity heterogeneity,
temporal connectivity, and attractiveness shape epidemic thresholds and
endemic prevalence~\cite{starnini_temporal_2014, zino_continuous-time_2016, pozzana_epidemic_2017}, and
developed strategies for controlling the coupled network and spreading dynamics
\cite{liu_controlling_2014}. Adaptive extensions allow infection, risk perception,
quarantine, or preventive behavior to change node activity and contact formation~\cite{rizzo_effect_2014, ogura_optimal_2019, mancastroppa_active_2020}. Li \emph{et al.} allowed prevalence to alter activity in a higher-order
activity-driven network and found two bistable regimes \cite{li_activity_2026}.
In these models, activity changes the temporal network, whereas in our model it changes
transmission over a fixed network.

Multistability also arises through mechanisms unrelated to activity-driven
contact formation. Susceptible nodes may rewire away from infected neighbours
\cite{gross_epidemic_2006}, recovery may depend on resources generated by the
healthy population \cite{bottcher_disease-induced_2015}, and group transmission may
reinforce contagion through higher-order interactions
\cite{iacopini_simplicial_2019}. These studies show that multistability can arise
when network adaptation, recovery feedback, or higher-order interactions make
transmission depend nonlinearly on the spreading state. Our model provides a
related mechanism in which transmission is modulated indirectly through
feedback with a coevolving dynamical state on the same network.

\subsection{Implications for neurodegenerative disease}

The main motivation for studying this coupled system is neurodegenerative
disease. Increased neuronal activity can promote tau release and propagation in Alzheimer's disease
\cite{pooler_physiological_2013, wu_neuronal_2016} and can modulate
$\alpha$-synuclein spreading \cite{wu_neuronal_2020} in Parkinson's disease, providing examples of a
positive activity-to-spreading coupling. Evidence that greater neuronal
activity instead suppresses pathological protein spreading, which would motivate a negative
activity-to-spreading coupling, is less established. By contrast, either sign
appears plausible for the spreading-to-activity coupling. Amyloid-$\beta$ has been
associated with neuronal hyperactivity
\cite{busche_clusters_2008, busche_critical_2012}, whereas tau can impair neuronal
circuits and suppress activity~\cite{busche_tau_2019, harris_tipping_2020}. The response to amyloid-$\beta$ may also
be nonmonotone: low picomolar concentrations potentiate synaptic transmission,
whereas higher nanomolar concentrations suppress it~\cite{puzzo_picomolar_2008, palop_amyloid-induced_2010}. An inverted-U, quadratic
pathology-to-activity coupling may therefore be biologically relevant. Combined
with a positive activity-to-spreading coupling, its rising branch reinforces
spreading at low pathology, whereas its falling branch opposes spreading at
higher pathology. Our theory predicts that sufficiently strong reinforcement
at low pathology can create a finite-amplitude invasion boundary and
disease-free--endemic bistability, while the later reversal limits further
reinforcement. Thus, the spreading-to-activity coupling may depend on the
protein species, disease stage, and amount of pathology.

Amyloid-associated hyperactivity and activity-dependent tau spreading suggest a
specific mechanism: amyloid accumulation may first raise neuronal activity,
which could then facilitate the later spread of tau.
Longitudinal human studies indicate that amyloid-$\beta$ accumulation precedes
the widespread neocortical accumulation of tau, and that an earlier rise in
amyloid is associated with subsequent tau accumulation~\cite{hanseeuw_association_2019, mattsson-carlgren__2020, sanchez_longitudinal_2021}. Amyloid-associated hyperactivity may therefore be
present before much of the later tau propagation. Since higher neuronal
activity promotes tau release and propagation
\cite{pooler_physiological_2013, wu_neuronal_2016}, the coupled SIS-QIF simulations presented herein suggest two
possible tipping mechanisms. First, amyloid-induced hyperactivity could push tau
pathology across a finite-amplitude invasion boundary, producing an abrupt
transition from health to widespread pathological propagation. Second, if tau is already
present at a low level, the same feedback could instead move the system from a
low- to a high-tau state through endemic--endemic bistability. Tau pathology can
indeed occur with little or no amyloid-$\beta$ in primary age-related tauopathy
(PART)~\cite{crary_primary_2014, josephs_tau_2017}. The mechanism described here may
explain how such tau pathology could become much more extensive after
amyloid-$\beta$ accumulates. The nonmonotone, approximately quadratic
effect of amyloid-$\beta$ on activity makes such state-dependent feedback
particularly plausible. Amyloid-$\beta$ may therefore make the brain more
amenable to subsequent tau propagation through its effect on neuronal
activity, consistent with recent human imaging findings
\cite{alexandersen_neuronal_2026}.

The spiking QIF--SIS system is bistable, and we show that external input to the
neuronal dynamics can move it between low- and high-pathology states. The change
persists after the input is removed. Neuronal stimulation has also altered
protein pathology experimentally. In mouse models, 40-Hz optogenetic or visual
stimulation reduced amyloid-$\beta$ levels \cite{iaccarino_gamma_2016}, while
auditory stimulation reduced amyloid and phosphorylated tau, and combined
auditory and visual stimulation reduced amyloid more widely
\cite{martorell_multi-sensory_2019}. These interventions may also act through
mechanisms which are not represented here. Nonetheless, our model suggests
that suitably chosen neuronal stimulation could, in principle, prevent or halt
pathological protein propagation by acting through activity--pathology
feedback.

\subsection{Limitations and outlook}

Several limitations should be noted. The center-manifold calculation uses a
node-level mean-field approximation and therefore neglects correlations between
infection states. Such correlations are especially important on sparse or
clustered networks and may shift both the invasion threshold and the saddle-node bifurcations. The calculation also assumes that disease-free activity approaches a
stable fixed point. If activity is periodic or irregular, the growth of a small
infection depends on the entire time-varying activity trajectory rather than on
a single activity value. Moreover,
the exclusion of periodic orbits and the global sign rule for multistability assume homogeneous dynamics on regular graphs. Their
conclusions may change on irregular or directed networks, with sign-changing transmission multipliers, feedback coupling that depends on both source- and
target-node states, or delays. Such extensions may support multistability with opposite-sign feedback, periodic orbits, and other dynamics outside the homogeneous reduction. Finally,
because the disease-free state of the finite stochastic system is absorbing,
the deterministic endemic branches represent long-lived quasistationary regimes
whose switching and extinction times may depend on network size. The minimal
examples and QIF models nevertheless reproduce the predicted multistable dynamics under
the assumptions studied here.

Taken together, these results suggest that neuronal activity may be more than an accelerator of pathology propagation. Through its feedback with pathology, activity can
reshape the dynamical landscape of disease, creating tipping points,
hysteresis, and multiple stable pathology levels. Transient changes in activity
may therefore produce lasting changes in disease progression. Applying this
framework to longitudinal multimodal neuroimaging may reveal signatures of such
tipping points in human disease and help determine whether targeted modulation
of activity can redirect pathology toward a lower-spreading state.

\bibliographystyle{unsrt}
\bibliography{references}

\clearpage
\appendix

\section{Mean-field closure, center-manifold reduction, and numerical continuation}
\label{app:continuous_computations}

\subsection{Markov generator and first-moment equation}

Let $X=(X_1,\ldots,X_N)^\top\in\{0,1\}^N$.
For $X_i=0$, let $X^{i,+}$ denote the state obtained by replacing
$X_i$ by one. For $X_i=1$, let $X^{i,-}$ denote the state obtained by
replacing $X_i$ by zero.
To distinguish the random microscopic activity from its deterministic
mean, write $U$ for the continuous component of the joint stochastic process
and define $u(t)=\mathbb E[U(t)]$. Between infection and clearance jumps,
$U$ obeys $\dot U=\mathcal G(U,X)$.

The generator of the joint piecewise-deterministic Markov process is
\begin{align}
\mathcal L\varphi(X,U)
={}&
\sum_{i=1}^N
(1-X_i)\lambda_i(X,U)
\left[
\varphi(X^{i,+},U)-\varphi(X,U)
\right]
\notag\\
&+
\sum_{i=1}^N
\zeta X_i
\left[
\varphi(X^{i,-},U)-\varphi(X,U)
\right]
\notag\\
&+
\nabla_U\varphi(X,U)^\top
\mathcal G(U,X),
\label{eq:joint_generator}
\end{align}
where $\lambda_i(X,U)=\beta\sum_{j=1}^N W_{ij}\Phi(U_j)X_j$.

Choose $\varphi(X,U)=X_i$.
The continuous activity term vanishes because $X_i$ does not depend
on $U$. The generator gives
$\mathcal LX_i=(1-X_i)\lambda_i(X,U)-\zeta X_i$.
Dynkin's formula therefore yields
\begin{align}
\frac{d}{dt}\mathbb E[X_i]
={}&
\beta
\sum_{j=1}^N
W_{ij}
\mathbb E
\left[
(1-X_i)
\Phi(U_j)
X_j
\right]
\notag\\
&-
\zeta\mathbb E[X_i].
\label{eq:exact_first_moment}
\end{align}

The equation is not closed because it contains joint infection and
activity moments. The node-level mean-field closure makes two distinct
approximations. First, it factorizes the mixed moment as
\[
\mathbb E
\left[
(1-X_i)
\Phi(U_j)
X_j
\right]
\approx
(1-x_i)\Phi(u_j)x_j,
\]
where $x_i=\mathbb E[X_i]$ and $u_j=\mathbb E[U_j]$. This closure neglects
correlations between $X_i$, $X_j$, and $U_j$, and also makes the nonlinear
activity approximation
\[
\mathbb E[\Phi(U_j)]
\approx
\Phi(\mathbb E[U_j])
=
\Phi(u_j),
\]
which is exact for affine $\Phi$ but not in general.
Substitution into~\eqref{eq:exact_first_moment} gives
\[
\dot x_i
=
\beta(1-x_i)
\sum_{j=1}^N
W_{ij}\Phi(u_j)x_j
-
\zeta x_i.
\]
Writing this equation for all nodes gives
\[
\dot x
=
\beta
(\mathbf 1-x)
\odot
W
\diag(\Phi(u))x
-
\zeta x.
\]

Second, applying the generator to the activity coordinates gives the exact
first-moment relation
\[
\frac{d}{dt}\mathbb E[U]
=
\mathbb E[\mathcal G(U,X)].
\]
The deterministic activity equation $\dot u=\mathcal G(u,x)$ therefore
assumes
\[
\mathbb E[\mathcal G(U,X)]
\approx
\mathcal G(u,x).
\]
Because $u=\mathbb E[U]$ and $x=\mathbb E[X]$, this identity is exact when
$\mathcal G$ is affine in $(U,X)$. For nonlinear $\mathcal G$, it is a
mean-field approximation justified when fluctuations and correlations are
sufficiently weak. Together, these two assumptions close the deterministic
system~\eqref{eq:full_continuous_system}.

\subsection{Center-manifold reduction at the epidemic threshold}
\label{app:center_manifold_reduction}

Set \(u=a+y\), where \(\mathcal G(a,\mathbf 0)=\mathbf 0\).  The shifted
deterministic system is
\[
\begin{aligned}
\dot x
&=
\beta(\mathbf 1-x)\odot
W\diag\bigl(\Phi(a+y)\bigr)x-\zeta x,
\\
\dot y
&=
\mathcal G(a+y,x).
\end{aligned}
\]
Let
\[
A=\diag(\Phi(a)),
\qquad
J_A=D_u\mathcal G(a,\mathbf 0),
\qquad
C=D_x\mathcal G(a,\mathbf 0).
\]
The disease-free equilibrium is \((x,y)=(\mathbf 0,\mathbf 0)\).  Since
\(J_A\) is stable, the transcritical bifurcation occurs at
\[
\beta_c=\frac{\zeta}{\rho_0},
\qquad
\rho_0=\rho(WA).
\]
Let \(r_0\) and \(\ell_0\) be the right and left Perron eigenvectors of
\(WA\), normalized by \(\ell_0^\top r_0=1\), and define
\[
L_c=\beta_cWA-\zeta I,
\qquad
h=-J_A^{-1}Cr_0.
\]
For the full state \(Z=(x^\top,y^\top)^\top\), the right and left nullvectors
of the threshold Jacobian are
\[
v=
\begin{pmatrix}r_0\\h\end{pmatrix},
\qquad
q=
\begin{pmatrix}\ell_0\\\mathbf 0\end{pmatrix},
\qquad
q^\top v=1.
\]

Let \(\mu=\beta-\beta_c\), write the right-hand side as
\(\dot Z=\mathcal F(Z;\beta_c+\mu)\), and choose the center coordinate
\[
z=q^\top Z=\ell_0^\top x.
\]
The center manifold theorem gives a locally invariant family
\(Z=\mathcal H(z,\mu)=zv+O(z^2,\mu z)\).  At the threshold we write its
quadratic approximation as
\[
\mathcal H(z,0)
=
\begin{pmatrix}
zr_0+z^2\xi\\
zh+z^2\psi
\end{pmatrix}
+O(z^3),
\qquad
\ell_0^\top\xi=0.
\]
Here \(\xi\) and \(\psi\) are the quadratic corrections to the infection and
activity components, respectively.  The reduced equation has the form
\[
\dot z
=
\rho_0\mu z+c_2z^2+c_3z^3
+O(\mu^2z,\mu z^2,z^4).
\]
The center-manifold invariance equation is
\[
D_z\mathcal H(z,\mu)\dot z
=
\mathcal F\bigl(\mathcal H(z,\mu);\beta_c+\mu\bigr).
\]

To determine the coefficients in the reduced equation, we substitute the
center-manifold approximation into the invariance equation.  For this purpose,
define
\[
\mathcal D_1[w]
=
\diag\bigl(\Phi'(a)\odot w\bigr),
\qquad
\mathcal D_2[w_1,w_2]
=
\diag\bigl(\Phi''(a)\odot w_1\odot w_2\bigr),
\]
and
\[
\mathcal Q_A(y,x)
=
\frac12D^2\mathcal G(a,\mathbf 0)
\left[
\begin{pmatrix}y\\x\end{pmatrix},
\begin{pmatrix}y\\x\end{pmatrix}
\right].
\]
Expanding the original system at \(\beta=\beta_c\) gives
\[
\begin{aligned}
\dot x
={}&
L_cx
-\beta_c x\odot WAx
+\beta_c W\mathcal D_1[y]x
\\
&-
\beta_c x\odot W\mathcal D_1[y]x
+\frac{\beta_c}{2}W\mathcal D_2[y,y]x
+O\!\left(\norm{(x,y)}^4\right),
\\
\dot y
={}&
J_Ay+Cx+\mathcal Q_A(y,x)
+O\!\left(\norm{(x,y)}^3\right).
\end{aligned}
\]
We now match powers of \(z\) and \(\mu\).  At order
\(\mu z\), projection with \(q^\top\) gives
\[
\ell_0^\top WAr_0=\rho_0.
\]
At quadratic order, let
\[
n_2
=
-\zeta\,r_0\odot r_0
+\beta_cW\mathcal D_1[h]r_0
\]
be the coefficient of \(z^2\) in the infection equation along the critical
direction.  Projection with \(\ell_0^\top\) gives
\[
\boxed{
c_2
=
-\zeta\,\ell_0^\top(r_0\odot r_0)
+\beta_c\ell_0^\top W\mathcal D_1[h]r_0.
}
\]

Let \(\mathcal S=\{w\in\mathbb R^N:\ell_0^\top w=0\}\), and denote the
inverse of the restriction of \(L_c\) to \(\mathcal S\) by
\(L_{c,\mathcal S}^{-1}\).  This inverse exists because the zero eigenvalue
of \(L_c\) is simple.  Since
\(\ell_0^\top(c_2r_0-n_2)=0\), the quadratic corrections are
\[
\boxed{
\begin{aligned}
\xi
&=
L_{c,\mathcal S}^{-1}(c_2r_0-n_2),
\\
\psi
&=
J_A^{-1}
\left[c_2h-C\xi-\mathcal Q_A(h,r_0)\right].
\end{aligned}
}
\]

It remains only to match the cubic terms.  After substituting the known
\(\xi\) and \(\psi\), the cubic terms in the
infection equation that contribute after projection are
\[
\begin{aligned}
n_3
={}&
-\beta_c
\left[
r_0\odot WA\xi
+\xi\odot WAr_0
\right]
\\
&+
\beta_c W
\left[
\mathcal D_1[h]\xi
+\mathcal D_1[\psi]r_0
+\frac12\mathcal D_2[h,h]r_0
\right]
\\
&-
\beta_c r_0\odot W\mathcal D_1[h]r_0.
\end{aligned}
\]
Projecting the matched cubic equation with \(\ell_0^\top\) and using
\(\ell_0^\top\xi=0\) gives
\[
\boxed{c_3=\ell_0^\top n_3,}
\]
which is Eq.~\eqref{eq:c3_explicit} in the main text and completes the cubic
center-manifold reduction.

The coefficients \(c_2\) and \(c_3\) are used for the center-manifold predictions in
Figure~\ref{fig:irregular_network_validation}(a,b).  Panel (c) also uses
\(c_4\).  We obtain it numerically by first determining the cubic
center-manifold correction from the unprojected \(z^3\) terms and then matching
the \(z^4\) terms in the same invariance equation.  We omit this standard but
lengthy fourth-order calculation.

\subsection{Numerical continuation and stability for Figure 1}
\label{app:irregular_network_numerics}

For Figure~\ref{fig:irregular_network_validation}, we use a fixed undirected
Erd\H{o}s--R\'enyi realization \(G(20,0.2)\) and normalize its adjacency matrix
so that \(\rho(W)=1\).  We set
\(\zeta=1\) and use, for panels (a--c), respectively,
\[
\Phi_1(u)=1+\frac{u}{2},
\qquad
\Phi_2(u)=(1+u)^2,
\qquad
\Phi_3(u)=1+\frac{3u}{5}+\frac{5u^2}{2}.
\]
For each \(k=1,2,3\), the full system is
\[
\dot x
=
\beta(\mathbf 1-x)\odot W\diag\bigl(\Phi_k(u)\bigr)x-x,
\qquad
\dot u=-u+x.
\]
Thus every equilibrium satisfies \(u=x\), and \(\beta_c=1\).

We parameterize each endemic branch by \(z=\ell_0^\top x>0\).  For each
coupling function \(\Phi_k\) and each prescribed \(z\), we solve the \(N+1\)
equations
\[
\begin{pmatrix}
\beta(\mathbf 1-x)\odot W\diag(\Phi_k(x))x-x\\
\ell_0^\top x-z
\end{pmatrix}
=
\mathbf 0
\]
for \((x,\beta)\).  The continuation starts near the threshold with
\(z_0=2\times10^{-5}\), using \(x=z_0r_0\) and \(\beta=\beta_c\) as the
initial guess.  For each \(k\), we use 700 equally spaced values of \(z\) from
\(z_0\) to \(z_{\max,k}=0.55,0.78,0.62\), respectively.  At each subsequent
value, the preceding equilibrium, with its infection vector rescaled by the
ratio of successive \(z\)-values, initializes a Levenberg--Marquardt solve of
the augmented equations above.  The solver tolerances are \(10^{-12}\).  This
continuation in \(z\) passes through folds in \(\beta\); the disease-free
branch is added separately.

The black dashed comparison curves are evaluated independently of the
center-manifold reduced equation, using \(c_2,c_3\) in panels (a,b) and
\(c_2,c_3,c_4\) in panel (c).

At every computed equilibrium, we evaluate the full \(2N\)-dimensional
Jacobian.  A branch is classified as stable when all eigenvalues have negative
real parts and unstable otherwise; these classifications give the solid and
dashed curves in Figure~\ref{fig:irregular_network_validation}(a--c).  The
trajectories in panels (d--f) are direct integrations of the full deterministic
system from the lower and higher initial conditions specified in the figure
caption.

\section{Nonlinear feedback on regular graphs}
\label{app:regular_polynomial_feedback}

\subsection{Exclusion of periodic orbits}
\label{app:regular_periodic_exclusion}

For the homogeneous system~\eqref{eq:regular_polynomial_system}, write
\[
f(x,y)
=
x\left[
\beta\rho(1-x)\Phi(a+y)-\zeta
\right],
\qquad
g(x,y)
=
-y+H(x).
\]
On the half-plane $x>0$, choose the Dulac function $B(x,y)=1/x$. Then
\begin{align*}
\frac{\partial}{\partial x}(Bf)
+
\frac{\partial}{\partial y}(Bg)
&=
\frac{\partial}{\partial x}
\left[
\beta\rho(1-x)\Phi(a+y)-\zeta
\right]
+
\frac{\partial}{\partial y}
\left[
\frac{-y+H(x)}{x}
\right]
\\
&=
-\beta\rho\Phi(a+y)-\frac1x
<0.
\end{align*}
The Bendixson--Dulac criterion therefore rules out a periodic orbit in any
simply connected physical region with $x>0$ and $\Phi>0$.  Here $\Phi>0$
means that activity may strengthen or weaken an
existing transmission edge without reversing its sign, which is a basic
assumption for the spreading processes considered here.  If an application
allows activity to drive $\Phi$ through zero and flip effective edge signs,
thereby changing the signed network structure, the divergence need not remain
negative and this argument no longer rules out periodic orbits.

%
%

\subsection{Exact polynomial examples and numerical verification}
\label{app:regular_exact_examples}

The following are the polynomial transmission functions used in
Figure~\ref{fig:regular_graph_multistability}.  We set $a=0$,
$\rho=\zeta=1$, and $H(x)=x$ in
Eq.~\eqref{eq:regular_polynomial_system}.  At an endemic equilibrium $y=x$.
Writing $\alpha:=\zeta/(\beta\rho)=1/\beta$, the equilibrium condition is
$(1-x)\Phi(x)-\alpha=0$.
The factorizations below therefore display all endemic infection levels for
the corresponding values of $\alpha$.

For the quadratic example, take
\[
\Phi_2(y)
=
1-
\frac{15}{7}y
+
\frac{50}{7}y^2,
\qquad
\alpha_2
=
\frac{27}{35}.
\]
Then
\begin{equation}
\label{eq:regular_quadratic_factorization}
(1-x)\Phi_2(x)-\alpha_2
=
-\frac{50}{7}
\left(x-\frac1{10}\right)
\left(x-\frac25\right)
\left(x-\frac45\right).
\end{equation}

For the cubic example, take
\[
\Phi_3(y)
=
1+12y-
\frac{340}{9}y^2
+
\frac{400}{9}y^3,
\qquad
\alpha_3
=
\frac{42}{25}.
\]
The exact factorization is
\begin{equation}
\label{eq:regular_cubic_factorization}
\begin{aligned}
(1-x)\Phi_3(x)-\alpha_3
=
-\frac{400}{9}
&\left(x-\frac1{10}\right)
\left(x-\frac3{10}\right)
\\
&\times
\left(x-\frac35\right)
\left(x-\frac{17}{20}\right).
\end{aligned}
\end{equation}

For the quartic example, take
\begin{equation}
\label{eq:regular_quartic_transmission}
\begin{aligned}
\Phi_4(y)
={}&
1-
\frac{14070}{1079}y
+
\frac{72200}{1079}y^2
\\
&-
\frac{132000}{1079}y^3
+
\frac{80000}{1079}y^4,
\qquad
\alpha_4
=
\frac{243}{1079}.
\end{aligned}
\end{equation}
Then
\begin{equation}
\label{eq:regular_quartic_factorization}
\begin{aligned}
(1-x)\Phi_4(x)-\alpha_4
=
-\frac{80000}{1079}
&\left(x-\frac1{10}\right)
\left(x-\frac14\right)
\left(x-\frac{11}{20}\right)
\\
&\times
\left(x-\frac45\right)
\left(x-\frac{19}{20}\right).
\end{aligned}
\end{equation}

All three transmission multipliers are positive for $0\leq y\leq1$, so the
examples satisfy the physical assumption $\Phi>0$.  For each polynomial, we
numerically computed the equilibria and their stability and simulated
Eq.~\eqref{eq:regular_polynomial_system} using an adaptive eighth-order
Runge--Kutta method.

\section{Monostability of the homogeneous QIF system}
\label{app:qif_monostability}

At fixed homogeneous infection $x$, setting the two QIF population equations
to zero gives
\begin{equation}
\label{eq:qif_stationary_equation}
Q(r;x)
=
\frac{\Delta^2}{4\pi^2r^2}
+\bar\eta+H(x)+(J_{\mathrm{loc}}+K)r-\pi^2r^2
=0,
\qquad
v=-\frac{\Delta}{2\pi r}.
\end{equation}
Here we used $\sum_jW_{ij}r=r$ on the homogeneous regular graph. For
$J_{\mathrm{loc}}+K\leq0$,
\begin{equation}
\label{eq:qif_stationary_monotonicity}
\partial_r Q
=
-\frac{\Delta^2}{2\pi^2r^3}+J_{\mathrm{loc}}+K-2\pi^2r
<0
\end{equation}
for every $r>0$. Moreover, $Q(r;x)$ tends to $+\infty$ as
$r\to0^+$ and to $-\infty$ as $r\to\infty$. It therefore has exactly one
positive root $r_*(x)$ for every fixed $x$. At the corresponding equilibrium,
the neuronal Jacobian satisfies
\[
\operatorname{tr}J_*
=\frac{4v_*}{\varepsilon}<0,
\qquad
\det J_*
=-\frac{2r_*}{\varepsilon^2}\partial_rQ(r_*;x)>0,
\]
so the equilibrium is locally asymptotically stable. Moreover, the Dulac
function $B(r,v)=r^{-2}$ gives
\[
\partial_r(B\dot r)+\partial_v(B\dot v)
=-\frac{2\Delta}{\varepsilon\pi r^3}<0,
\]
excluding periodic orbits in the physical half-plane $r>0$. Finally,
differentiating the stationary equation gives
$r_*'(x)=-H'(x)/\partial_rQ$, which has the same sign as $H'(x)$ and hence,
for $H(x)=\kappa S_x(x)$ with increasing $S_x$, the same sign as $\kappa$.

\section{Simulation of the microscopic QIF--SIS system}
\label{app:microscopic_simulation}

\subsection{Numerical method}
\label{app:microscopic_numerical_method}

Each node carries a circular theta-neuron phase $\theta_i$, an exponentially
filtered spike variable $s_i$, and a binary SIS state $X_i\in\{0,1\}$.
Infection changes the neuronal current by $\kappa X_i$, while the filtered
firing of an infected source changes its outward transmission through
$1+\delta S_r(s_i)$. We simulate these coupled processes using a split-step
scheme of the type used for coupled neuronal ODE and Markov-jump
models. Over each short interval, we first advance
the neuronal variables with infection fixed and then sample the SIS jumps with
the updated filtered firing fixed.

We use the fixed splitting step $\Delta t=0.002$. Let $X^n$, $\theta^n$, and
$s^n$ denote the states at the beginning of $[t_n,t_n+\Delta t]$. Holding
$X^n$ and $s^n$ fixed, we calculate the neuronal current
\[
I_i^n
=
\eta_i+\kappa X_i^n
+J\sum_jW_{ij}s_j^n
+I_i^{\rm ext}(t_n).
\]
Writing $f(\theta,I)=1-\cos\theta+(1+\cos\theta)I$, we advance the phase by
the explicit midpoint rule
\[
\theta_i^{n+1}
=
\theta_i^n
+\frac{\Delta t}{\varepsilon}
f\!\left(
\theta_i^n
+\frac{\Delta t}{2\varepsilon}f(\theta_i^n,I_i^n),
I_i^n
\right).
\]
We set $N_i^n=1$ if this midpoint step carries the phase of neuron $i$ to or
beyond the spike value $\theta_i=\pi$, and $N_i^n=0$ otherwise. The filtered
firing is then updated as
\[
s_i^{n+1}
=
\exp\!\left(-\frac{\Delta t}{\tau_s}\right)s_i^n
+\frac{\varepsilon}{\tau_s}N_i^n.
\]

We then keep $s^{n+1}$ fixed and simulate the SIS process over the same
interval using the direct Gillespie method~\cite{gillespie_exact_1977}. For a
susceptible node, the infection rate is
\[
\lambda_i(X;s^{n+1})
=
\frac{\beta}{\tau_X}
\sum_jW_{ij}\bigl(1+\delta S_r(s_j^{n+1})\bigr)X_j.
\]
Thus, for the current binary state $X$, the total event rate at node $i$ is
\[
a_i(X;s^{n+1})
=
(1-X_i)\lambda_i(X;s^{n+1})
+X_i\frac{\zeta}{\tau_X},
\]
where the first term is the infection rate when node $i$ is susceptible and
the second is the recovery rate when it is infected. Let $a_0=\sum_i a_i$.
We draw a waiting time from an exponential distribution with rate $a_0$. If
it exceeds the time remaining in the interval, no further event occurs.
Otherwise, node $i$ is selected with probability $a_i/a_0$ and its state is
switched. We then recompute the SIS event rates for the updated binary state
and repeat only this Gillespie event calculation until the interval is
exhausted; $\theta^{n+1}$ and $s^{n+1}$ remain fixed throughout this SIS
substep. The resulting state is $X^{n+1}$. The next splitting interval starts
from $(X^{n+1},\theta^{n+1},s^{n+1})$, and the same neuronal and SIS substeps
are repeated.

The neuronal excitabilities are the deterministic Lorentzian quantiles
\[
\eta_i
=
\bar\eta
+\Delta\tan\!\left[
\pi\left(\frac{i}{N+1}-\frac12\right)
\right],
\qquad i=1,\ldots,N.
\]
These are values of the inverse Lorentzian cumulative distribution function
at the evenly spaced probabilities $i/(N+1)$. Choosing them deterministically
rather than drawing them randomly makes the finite network represent the
intended Lorentzian distribution evenly and eliminates sample-to-sample
variation in the excitability values. We randomly permute this fixed set
across the nodes to avoid correlations with network structure. Initially,
each $\theta_i$ is drawn uniformly from $[-\pi,\pi]$ and
$s_i=0.01$. The initially infected nodes are selected uniformly at random,
with their number specified separately for each experiment.

The spatial graph is generated once, with node positions sampled independently
from a uniform distribution over the unit square and edges between nodes
separated by at most $0.155$. For
Figures~\ref{fig:microscopic_qif_stochastic_regimes} and
\ref{fig:microscopic_qif_subthreshold_bistability}, the two unforced
trajectories are integrated to $t=1200$ and the input-driven trajectory to
$t=5500$, with all trajectories sampled every $0.5$.

\subsection{Control with opposite-sign coupling}
\label{app:microscopic_opposite_sign_control}

\begin{figure}[!htb]
\centering
\includegraphics[width=0.86\textwidth]{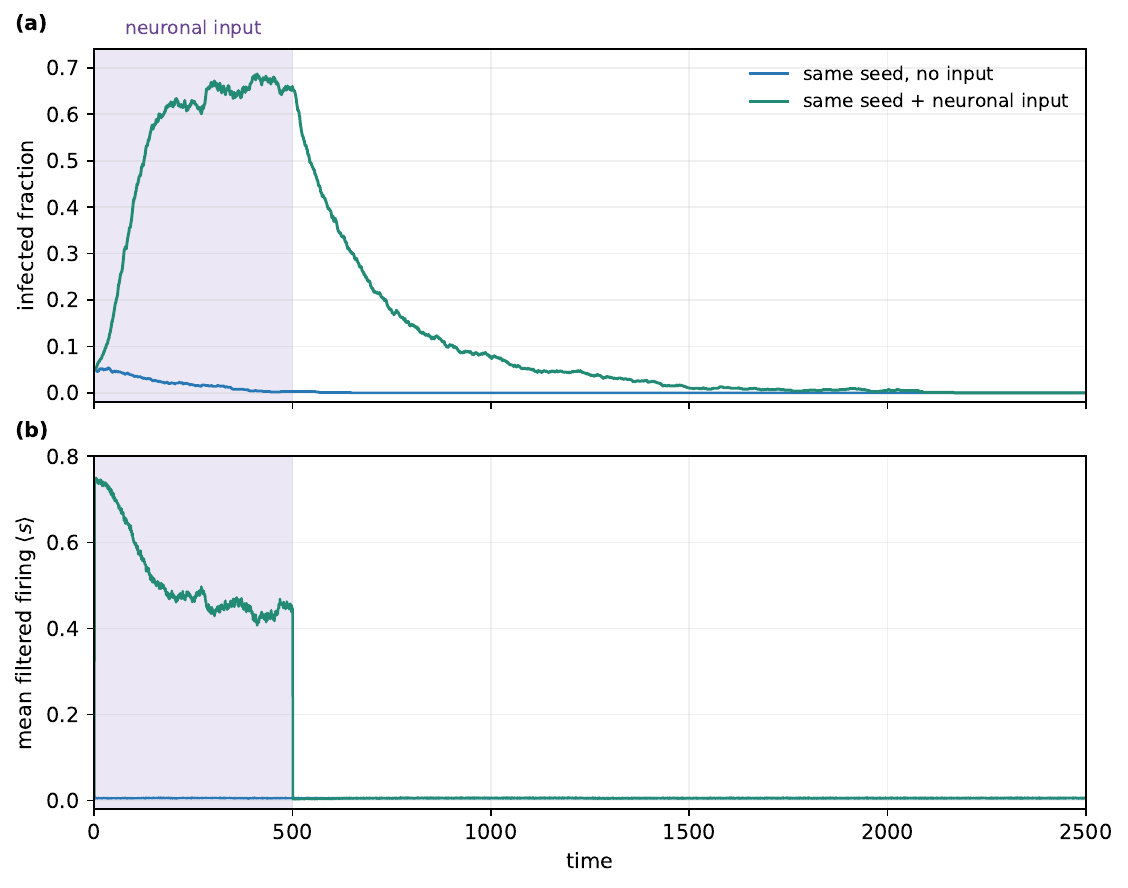}
\caption{Opposite-sign control at $\beta=0.76$. (a) Infection decays after
the neuronal input is removed. (b) Mean filtered firing returns to baseline.
The graph, 50-node infection seed, and remaining parameters are as in
Figure~\ref{fig:microscopic_qif_subthreshold_bistability}, except
$\kappa=-3$. The input has amplitude $5$ and acts over $0\leq t<500$.}
\label{fig:microscopic_qif_opposite_sign_control}
\end{figure}

Figure~\ref{fig:microscopic_qif_subthreshold_bistability} showed that, at
$\beta=0.76$ and with same-sign feedback $(\kappa,\delta)=(3,12)$, a transient
positive neuronal input can drive a 50-node infection seed into a high-endemic
state which persists after the input is removed. The following control tests
whether this persistence is merely slow stochastic decay. Recall that $\kappa$
determines how infection changes the neuronal current, whereas $\delta$
determines how neuronal firing changes transmission. In
Figure~\ref{fig:microscopic_qif_opposite_sign_control}, we repeat the experiment
with the same graph and stochastic initialization but reverse $\kappa$ from
$3$ to $-3$, while keeping $\delta=12$. Infection therefore suppresses firing,
whereas firing still promotes transmission, so the two coupling directions
have opposite signs. We apply the stronger input $I_i^{\rm ext}=5$ for
$0\leq t<500$, which raises the prevalence to $0.687$, comparable with the
high-endemic range in the same-sign example. Once the input is removed, firing
returns to its baseline level and infection decays to absorption at $t=2167$.

\end{document}